\documentclass[prl,floatfix,amsmath,superscriptaddress,twocolumn]{revtex4}

\usepackage{amssymb}
\usepackage{graphicx}
\usepackage{graphics}
\usepackage{amsmath}
\usepackage{mathtools}
\usepackage{amsthm}
\usepackage{amsfonts}
\usepackage{url}
\usepackage{booktabs}
\usepackage{braket}
\usepackage{bm}
\usepackage{bbold}
\usepackage{float}
\usepackage{enumerate}
\usepackage{color}
\usepackage{hyperref}
\usepackage{multirow}
\usepackage{array}

\hypersetup{
	colorlinks=true,
	linkcolor=blue, 
}
\def\bi{\begin{itemize}}
\def\ei{\end{itemize}}
\def\bc{\begin{center}}
\def\ec{\end{center}}

\def\C{\hbox{$\mit I$\kern-.7em$\mit C$}}
\def\R{\hbox{$\mit I$\kern-.6em$\mit R$}}
\def\N{\hbox{$\mit I$\kern-.65em$\mit N$}}

\def\ket#1{|#1\rangle}
\newcommand{\one}{\mbox{$1 \hspace{-1.0mm}  {\bf l}$}}
\def\tr{\mathrm{tr}}
\def\ket#1{\left| #1\right>}

\newcommand{\abs}[1]{\left| #1 \right|}

\newtheorem{theorem}{Theorem}

\newtheorem{lemma}[theorem]{Lemma}

\newtheorem{observation}[theorem]{Observation}

\makeatletter
\renewcommand{\maketag@@@}[1]{\hbox{\m@th\normalsize\normalfont#1}}
\makeatother

 \makeatletter
\newcommand*\normalcr\@normalcr
\makeatother
 
\begin{document}

\author{D. Sauerwein}
\affiliation{Institute for Theoretical Physics, University of
Innsbruck, Innsbruck, Austria}

\author{K. Schwaiger}
\affiliation{Institute for Theoretical Physics, University of
Innsbruck, Innsbruck, Austria}

\author{M. Cuquet}
\affiliation{Institute for Theoretical Physics, University of
Innsbruck, Innsbruck, Austria}

\author{J.I. de Vicente}
\affiliation{Departamento de Matem\'aticas, Universidad Carlos III de
Madrid, Legan\'es (Madrid), Spain}

\author{B. Kraus}
\affiliation{Institute for Theoretical Physics, University of
Innsbruck, Innsbruck, Austria}

\title{The source and accessible entanglement of few-body systems}

\date{\today}

\begin{abstract}
Entanglement is the resource to overcome the natural limitations of spatially separated parties restricted to Local Operations assisted by Classical Communications 
(LOCC). Recently two new classes of operational entanglement measures, the source and the accessible entanglement, for arbitrary multipartite states have 
been introduced. Whereas the source entanglement measures from how many states the state of interest can be obtained
via LOCC, the accessible entanglement measures how many states can be reached via LOCC from the state at hand. We consider here pure bipartite as well as 
multipartite states and derive explicit formulae
for the source entanglement. Moreover, we obtain explicit formulae for a whole class of source entanglement measures that characterize the simplicity of
generating a given bipartite pure state via LOCC. Furthermore, we show how the accessible entanglement can be computed numerically. For generic four--qubit states we first derive the necessary and sufficient conditions for the existence of LOCC transformations among these states and then derive explicit formulae for their accessible and source entanglement.
\end{abstract}
\maketitle

\section{Introduction}

Entanglement is considered to be one of the characteristic traits of quantum mechanics. Besides its interest from the foundational point of view, 
it plays a key role in quantum information science, being a resource for most of its applications such as quantum communication and quantum 
computation \cite{Nie}. This has led to the development of entanglement theory \cite{HoHo08}, which has also recently brought new tools for the 
understanding of quantum many-body systems in condensed matter physics \cite{Faz}. The characterization (of different forms) of entanglement and its 
quantification play a central role in this theory. Ideally, entanglement measures should allow to operationally quantify how useful a state is for 
quantum information applications, how efficient protocols can be and provide quantitative means to link entanglement to other physical phenomena such as
phase transitions. Despite remarkable advances, there are still many issues that require a better understanding, particularly in the case of multipartite 
states and mixed states in general. There is a clear need for entanglement measures which are on the one hand operationally meaningful and on the other 
hand easy to compute.

The paradigm of local operations assisted by classical communication (LOCC) is of paramount importance in the theory of entanglement. LOCC constitutes the
most general class of transformations allowed by the rules of quantum mechanics to distant parties: each subsystem can undergo any form of local quantum 
dynamics (i.e. a completely positive map) at will and the parties' actions can be correlated through the use of classical communication. Thus, 
LOCC maps constitute all possible protocols for the manipulation of entangled states. Moreover, LOCC allows to formulate entanglement theory as a 
resource theory. This is because entanglement cannot be created by LOCC alone and it is, hence, a resource to overcome the limitations of parties 
restricted to this class of operations. The basic law of entanglement is that it cannot increase by LOCC and, therefore, an entanglement measure must 
be a function that is monotonic with respect to the ordering induced by this class of transformations. Unfortunately, this ordering is in general just 
partial (see e.g. \cite{Nielsen}) and many different functions that qualify as entanglement measures exist. For this reason, it is desirable to seek for 
some further operational meaning behind these functions in the context of LOCC in order to single out the most relevant entanglement measures.

The identification of the operational meaning of entanglement measures has been particularly successful in the case of pure bipartite entangled states. 
It has been shown that the entropy of entanglement represents the rate at which any such state can be reversibly transformed by LOCC in the asymptotic limit
of infinitely many copies into the maximally entangled state \cite{entent}. Here, the notion of maximal entanglement stems from LOCC transformations on
single copies: the maximally entangled bipartite state is the unique state that cannot be obtained from any other Local Unitary (LU)-inequivalent bipartite
state but allows to be transformed into any other state \cite{Nielsen}. Regrettably, the idea of reversible entanglement distillation cannot be extended to 
bipartite mixed states. The entanglement cost (i.e. the rate at which the state is obtained from the maximally entangled state) is in general different 
from the distillable entanglement (i.e. the rate at which the state is transformed to the maximally entangled state) \cite{irrev}. Furthermore, the computation of these measures is formidably difficult. There are other measures that boil down to the entanglement entropy for pure states but which are different for mixed states such as the relative entropy of entanglement \cite{ree} or the squashed entanglement \cite{ChWi}. However, despite interesting properties, they lack the interpretation in terms of LOCC-conversion rates and their exact computation is still very hard.

The situation is even worse in the multipartite case. There exists no unique maximally entangled state \cite{mes} and it is not clear to which set of states LOCC asymptotic rates have to be defined (the so-called MREGS (minimal reversible entanglement generating set) problem \cite{mregs}). Nevertheless, several entanglement measures have been proposed from the purely mathematical perspective of invariants such as the tangle \cite{tangle} to the ability to create bipartite entanglement among given cuts such as the localizable entanglement \cite{locEnt}. Besides this, many bipartite measures admit a generalization to multipartite states like distance-based measures such as the geometric measure of entanglement \cite{gme} to cite an example. However, all these measures are hard to compute and, more importantly, many of them do not have an interpretation in the context of the LOCC paradigm.

In order to try to close this gap, we have recently introduced in \cite{us} operational entanglement measures for general (pure or mixed) multipartite states
which are LOCC-meaningful. The basic idea is to focus on the capacity of a state for single-copy LOCC transformations instead of transformation rates in
the asymptotic regime. In more detail, we have introduced two classes of entanglement measures based on the \emph{accessible volume} and the
\emph{source volume} of a state. The first one quantifies the relative volume of inequivalent states that can be accessed from our state by LOCC while the
second one quantifies the relative volume of inequivalent states from which our state can be obtained by LOCC. Hence, these measures have a clear
operational meaning in the context of LOCC transformations. The larger the accessible volume is the more entangled (i.e. useful) a state should be:
the state is at least as powerful in applications as any state in the accessible set as any protocol achievable with the latter can also be achieved with
the former by converting it by LOCC to this accessible state first. On the other hand, the larger the source volume the less entangled a state should
be given that many inequivalent states can reach this state by LOCC. Notice then that these measures are completely general, valid for arbitrary states of
any dimension. Moreover, they can be generalized to classes of entanglement measures by considering different Hilbert spaces for the initial and the 
final state (see \cite{us} and Sec. \ref{sec:AccandSourc}). Furthermore, whenever the possible LOCC transformations are characterized, these measures can
be computed. Although characterizing LOCC transformations is notoriously difficult \cite{Chitambar}, this problem is experiencing some progress lately 
\cite{Nielsen,LOCC,mes}. Using these insights, in \cite{us} we have analyzed the 3-qubit pure-state case providing explicit formulae for these measures. Here, we further show the versatility of this approach by computing these measures for the bipartite pure-state case and the generic four-qubit pure-state case. In order to do so, we complete the analysis of \cite{mes} and characterize all possible LOCC transformations among the latter class of states.

The outline of the remainder of the paper is the following. 
In Sec. II we recall the definition and properties of the two classes of entanglement measures we introduced in \cite{us}.
In Sec III we consider general bipartite pure states and derive explicit formulae for the source entanglement and its generalizations. That is, 
we obtain explicit formulae for a whole class of operational entanglement measures that 
quantify how easy it is to generate a bipartite pure state from other bipartite quantum states via LOCC. These measures can be used to 
characterize, e.g.,  the entanglement contained in pure quantum states of two qubits or two qutrits. Moreover, we demonstrate how the accessible volume can be computed and illustrate the results by considering up to two four--level systems.
Sec. IV deals with the multipartite case, where we first derive the necessary and sufficient conditions for LOCC convertibility of generic four-qubit 
states and then compute the new entanglement measures.

\section{Accessible and source entanglement}
\label{sec:AccandSourc}
In this section we review the definition of the entanglement measures we have introduced in \cite{us}. Since local unitary (LU) transformations are reversible LOCC transformations, the entanglement of two states is equivalent if they are related by such a transformation. Hence, we consider the possible LOCC transformations among the LU-equivalence classes of states rather than states in general. That is, we always pick a unique representative state from each class. We say that a state $\rho$ can {\it reach} a state $\sigma$ and that $\sigma$ is {\it accessible} from $\rho$ if there exists a deterministic LOCC protocol which transforms $\rho$ into $\sigma$. For a given state, $\rho$, we denote by $M_a(\rho)$ the set of states which can be accessed from $\rho$ and by $M_s(\rho)$ the set of states which can reach $\rho$ (see Fig.\ 1). Let $\mu$ denote an arbitrary measure in the set of LU equivalence classes of states. Then, the \textit{source volume} of $\rho$ is defined by $V_s(\rho)=\mu[M_s(\rho)]$ and the \textit{accessible volume} by $V_a(\rho)=\mu[M_a(\rho)]$. As mentioned in the introduction, these quantities measure respectively the amount of inequivalent states that are not less useful and not more useful than the state at hand. Thus, the \textit{accessible entanglement} and the \textit{source entanglement} are defined by
\begin{equation}\label{eas}
E_{a}(\rho)=\frac{V_a(\rho)}{V_a^{sup}},\quad E_{s}(\rho)=1-\frac{V_s(\rho)}{V_s^{sup}},
\end{equation}
where $V_a^{sup}$ ($V_s^{sup}$) denote the supremum of the accessible (source) volume according to the measure $\mu$.

\begin{figure}[H]\label{VaVsfig}
   \centering
   \includegraphics[width=0.25\textwidth]{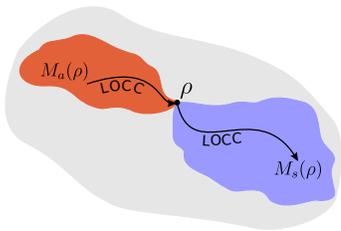}
   \caption{(color online). In this schematic figure the source set, $M_s(\rho)$, and the accessible set, $M_a(\rho)$, of the state $\rho$ are depicted. Any state in $M_s(\rho)$ can be transformed to $\rho$ via LOCC and $\rho$ can be transformed into any state in $M_a(\rho)$ via LOCC.}
   \label{VaVs}
   \end{figure}

A few remarks are in order. First, due to their operational meaning, it is easy to see that $E_a$ and $E_s$ are valid entanglement measures (see \cite{us}). That is, for any deterministic LOCC protocol $\Lambda$ it holds that \begin{equation}\label{monotonicity}
E_a(\Lambda(\rho))\leq E_a(\rho),\quad E_s(\Lambda(\rho))\leq E_s(\rho)
\end{equation}
for every state $\rho$.

Notice that the so-called entanglement monotones have been sometimes considered as entanglement quantifiers. These quantities must, in contrast to entanglement measures, fulfill that if different outcomes of the LOCC map $\Lambda$ acting on a state, $\Lambda_k(\rho)$, can be postselected, each of them occurring with probability $p_k$, then entanglement cannot increase on average, i.e. $\sum_kp_kE(\Lambda_k(\rho))\leq E(\rho)$. However, we follow here the definition of entanglement measures, for which $E(\Lambda(\rho))\leq E(\rho)$ for any LOCC map $\Lambda$ must hold. For a discussion justifying the suitability of this requirement over the averaged one, we refer the reader to Sec.\ XV.B.1 of \cite{HoHo08}.

Second, the choice of the mathematical measure $\mu$ to compute the volumes of the source and accessible sets is in principle arbitrary. That is,
any valid measure leads to an entanglement measure. Depending on the measure, one might obtain a different ordering of the states, as is common in case
of larger systems than those composed of two qubits \cite{KrCi00}. The choice of the measure can be fixed by the physical constraints of the problem at hand
or even out of mathematical convenience. Obviously however, one should always consider a fixed dimension of the states in the source set. This is because,
by imposing no constraint on the dimension, the size of this set would explode by considering source states of arbitrarily large dimension. A similar thing
happens for the accessible set as it would shrink to measure zero relative to states of sufficiently large dimension. Thus, one should always restrict to a
choice of measure which is supported on states of a given dimensionality. Note that, by changing this choice of dimension
(and/or the number of considered subsystems), one has the freedom to obtain different families of measures $\{E_s^k\}_{k\geq d}$ and $\{E_a^k\}_{k\leq d}$,
where $d$ is the effective dimension of the subsystems of the state at hand \cite{us}.

Another issue to take into account is that it can be
the case that one is interested in the relative volumes of states under LOCC transformations among particular classes of states. In this case one should
choose measures which are only supported on these classes. For instance, and this will always be the case in this article, one can consider only
transformations among pure states. Hence, the measure $\mu$ will only be supported on the set of LU-equivalence classes of pure states. This seems a
reasonable choice as the study of pure entanglement inside the full (mixed) state space might be too coarse grained given that pure states are of measure zero (e.g. $M_s$ would have measure zero for all pure two-qubit states on the full state space of two-qubit states given that it is not possible to convert via LOCC a non-pure two-qubit state into a pure entangled two-qubit state \cite{kent}). Another instance is the study of multipartite states where we will consider measures that are not supported on biseparable states. Finally, it might be even desirable to choose different mathematical measures for different subclasses. This might be the case when these subclasses are not LOCC related (thus, this does not compromise the validity of Eq.\ (\ref{monotonicity})) or when the different subclasses have source or accessible sets living in manifolds of different dimensionality (see \cite{us}).

\section{Bipartite systems}

\subsection{Preliminaries}

\subsubsection{LOCC transformations for bipartite systems}

Every pure state of a bipartite quantum system with Hilbert space $\mathcal{H} = \mathbb{C}^{d_1} \otimes \mathbb{C}^{d_2}$
can be (up to local unitaries) written as $\ket{\psi} \simeq_{LU} \sum_{i=1}^d \sqrt{\lambda_i} \ket{ii}$,
where $d = \min\{d_1,d_2\}$ and $\lambda_i \geq 0$ denote the Schmidt coefficients with $\sum_i \lambda_i = 1$. We denote by
$\lambda(\psi) = (\lambda_1, \ldots, \lambda_d) \in \R^d$
the Schmidt vector of $\ket{\psi}$ and will consider in the following w.l.o.g. two $d$-level systems. As can be easily seen
from the Schmidt decomposition given above, two $d$-level states, $\ket{\psi},\ket{\phi},$ are LU equivalent if and only if (iff)
$\lambda^\downarrow (\psi)=\lambda^\downarrow (\phi)$, where here and in the following $\lambda^\downarrow (\psi)\in \R^d$, with
 $\lambda^\downarrow(\psi)_{i} \geq \lambda^\downarrow(\psi)_{i+1} \geq 0$ denotes the sorted Schmidt vector of $\ket{\psi}$. In the context of
 LOCC transformations of pure bipartite states, the following functions of $x = (x_1,\ldots,x_d) \in \R^k$,
 \begin{align}
  E_k(x) := \sum_{i=1}^k x_i, \ k \in \{1,\ldots,d\},
 \end{align}
 play an important role. It was shown in \cite{Nielsen}
 that a state $\ket{\psi} \in \mathcal{H}$ can be transformed into $\ket{\phi} \in \mathcal{H}$ deterministically via
 LOCC iff $\lambda(\psi)$ is majorized by $\lambda(\phi)$, written
 $\lambda(\psi) \prec \lambda(\phi)$, i.e.
 \begin{align} \label{Ineq_Maj} E_k(\lambda^\downarrow(\psi)) \leq E_k(\lambda^\downarrow(\phi)) \ \forall k \in \{1,\ldots,d\},\end{align}
 with equality for $k=d$.

A direct consequence of this criterion
 is that the source and accessible set of $\ket{\psi} \in \mathcal{H}$ are given by
\begin{align}
 M_s(\psi) = \{\ket{\phi}\in {\cal H} \ \mbox{s.t.} \ \lambda (\phi) \prec \lambda (\psi)\}, \label{Eq_MsMastates1}\\
 M_a(\psi) = \{\ket{\phi}\in {\cal H} \ \mbox{s.t.} \ \lambda (\psi) \prec \lambda (\phi)\}. \label{Eq_MsMastates2}
\end{align}

As mentioned above, the volumes of these sets measure how easy it is to generate
a given state and how many states a given state can access respectively. As explained before, we quantify the set of LU-equivalence classes
as the amount of LU-equivalent states which can be used to reach a given state is not of relevance here. Due to the one-to-one correspondence between
the LU-equivalence classes of bipartite pure states and the sorted Schmidt vectors, we can associate to the sets given in Eqs. (\ref{Eq_MsMastates1}) and
(\ref{Eq_MsMastates2}) the following sets of sorted Schmidt vectors in $\R^d$:

\begin{align}
 \mathcal{M}_s(\psi) = \{\lambda^\downarrow \in \R^d \ \mbox{s.t.} \ \lambda^\downarrow \prec \lambda(\psi)\},\label{Eq_MsMa}\\
 \mathcal{M}_a(\psi) = \{\lambda^\downarrow \in \R^d \ \mbox{s.t.} \ \lambda(\psi) \prec \lambda^\downarrow\} \label{Eq_MsMa1}.
\end{align}
Unless otherwise stated, $d$ denotes the Schmidt rank of $|\psi\rangle$ and these sets are hence supported on states of 
the same dimensions as $|\psi\rangle$. Notice, however, that, as explained in the previous section, it is possible to consider source sets 
supported on states of larger dimension and accessible sets supported on states of smaller dimension. This gives the possibility to obtain different 
families of measures (see \cite{us} and Sec. \ref{sec:SourceVolume} below).

The sets given in Eqs. (\ref{Eq_MsMa})-(\ref{Eq_MsMa1}) are convex polytopes. In order to compute their volumes in the subsequent sections, we first review some of their properties.

\subsubsection{Convex polytopes}
We briefly recall some definitions and results concerning convex polytopes. The reader is referred to \cite{Ziegler}
for a comprehensive introduction into the study of convex polytopes. A closed halfspace in $\R^k$ is a set of the form
$H = \{x \in \R^k \ \mbox{s.t.} \ cx + c_0 \geq 0\}$, with $c \in \R^k, c_0 \in \R$.
We denote by $h(H)$ the hyperplane that fulfills the inequality with equality, i.e. $h(H):= \{x \in \R^k \ \mbox{s.t.} \ cx + c_0 = 0\}$ .
A well-known and important result for subsets $P \subset \R^k$ is the following equivalence. A subset $P \subset \R^k$ is the bounded intersection of a finite set of closed halfspaces, i.e.
\begin{align}
P = \{x \in \R^k \ \mbox{s.t.} \ Ax+b \geq 0\} \ \mbox{for} \ A \in \R^{m \times k}, b \in \R^m, \label{H-representation}
\end{align}
iff it is the convex hull of a finite point set, i.e.
\begin{align}
P = \text{conv}(V), \ \mbox{where} \ V = \{v_i \in \R^k\}.
\end{align}
The object $P$ is called a convex polytope and a representation of the form in Eq. (\ref{H-representation}) is an
$H$-representation of $P$. The latter way to define
the polytope is called $V$-representation. It is
unique if the set $\{v_i\}$ is minimal. This minimal set is called vertex set, $\text{vert}(P)$, of $P$. The dimension of $P$, $\text{dim}(P)$, is the
dimension of its affine hull, i.e. of
the smallest affine subspace of $\R^k$ that contains $P$. A face $F$ of $P$ is a set of the form $F = P \cap h(H)$ for some hyperplane $h(H)$. Here, $H \supset P$ is
a closed halfspace that contains $P$. The dimension of $F$ is again the dimension of its affine hull. Vertices are faces of dimension $0$, while
faces of dimension $1$ and $\text{dim}(P)-1$ are called edges and facets
of $P \subset \R^k$ respectively. Furthermore, a polytope with $\text{dim}(P) = k$ is called simple if every vertex is contained in only
$k$ facets. Note that this is the minimal number of facets a vector must be an element of in order to be a vertex. That is, every vertex of a simple polytope
is an element of only $k$ hyperplanes that are associated to closed halfspaces in a minimal $H$-representation, i.e. an $H$-representation of the
polytope that consists of the minimal number of intersecting halfspaces. Two vertices are called neighbors if they
are connected by an edge, i.e. if $k-1$ of the above mentioned hyperplanes coincide.
It is easy to see that a $k$-dimensional polytope is simple iff every vertex has $k$ neighbors
\footnote{It is well-known that a $k$-dimensional polytope is simple iff every vertex is adjacent to exactly $k$ edges (see e.g. \cite{Ziegler}).
 Each of these edges contains exactly one neighbor of the vertex.  This follows from the fact that, if it would not contain
any neighbor, the polytope would be unbounded. Whereas, if there were more
neighbors in this edge, one of them could be written as the convex combination of the initial vertex and the other neighbor. This would
 contradict the fact that the vertex set is minimal. Hence, a $k$-dimensional polytope is simple iff every vertex has $k$ neighbors.}.

Whereas sets such as $\mathcal{M}_s(\psi)$ have been studied in detail in the literature (see \cite{Ziegler} and references therein), and its vertices have
been computed \cite{Rado} (see Sec. \ref{sec:SourceVolume}), much less is known about the accessible set.
Note that not even the identification of the vertices of the accessible volume is straightforward.
In fact, even the number of vertices depends strongly on the state of interest, as we will explain in Sec. \ref{sec:Accessible}.
However, there exist algorithms that can compute the vertices, i.e. the minimal $V$-representation, of a convex polytope from a given $H$-representation.
They can also be used to compute the accessible volume, which we will use in Sec. \ref{sec:Accessible} to determine the accessible entanglement.

\subsection{Source volume}
\label{sec:SourceVolume}

We determine here the volume of the source set given in Eq. (\ref{Eq_MsMa}). In order to do so, we give a description of $\mathcal{M}_s(\psi)$ in terms of
convex geometry. Let us first note that the conditions in
Eq. (\ref{Ineq_Maj}) on the sorted Schmidt vectors are equivalent to the following set of inequalities on the unsorted Schmidt vectors

\begin{align}
\varOmega = \left \{E^{\sigma}_k(\lambda(\phi)) \leq E_k(\lambda^\downarrow(\psi)) \ \mbox{s.t.} \ k \in \{1,\ldots,d\},
\sigma \in \Sigma_d \right \}
\label{Ineq_Majunsort}
\end{align}
Here, $E^{\sigma}_k(\lambda):=E_k(P_{\sigma}\lambda)$, where $\sigma$ denotes an element of the permutation group $\Sigma_d$ of $d$ elements,
and $P_\sigma$ denotes the corresponding $d\times d$ permutation matrix, i.e.
$ P_{\sigma}\lambda = (\lambda_{\sigma(1)}, \lambda_{\sigma(2)},\ldots,\lambda_{\sigma(d)}) \equiv \lambda_{\sigma}$. This can be easily seen by noting
that for any $\lambda \in \R^d$ we have $E_k(P_{\sigma}\lambda) \leq E_k(\lambda^\downarrow)$ for any $k$ and any permutation $\sigma \in \Sigma_d$.
Hence, the majorization conditions given in Eq. (\ref{Ineq_Maj}) are fulfilled iff the conditions in Eq. (\ref{Ineq_Majunsort}) are.

For reasons that will become clear below, we consider the set of all Schmidt vectors that majorize $\lambda(\psi)$, not just the sorted ones, which is
given by

\begin{align}  \mathcal{M}^{LU}_s(\psi) &= \{\lambda \in \R^d \ \mbox{s.t.} \ \lambda \prec \lambda(\psi)\}. \end{align}

It is a well-known fact that $\mathcal{M}^{LU}_s(\psi)$ is a convex polytope whose vertices are given by the vectors $\lambda(\psi)_{\sigma}$,
with $\sigma\in \Sigma_d$ \cite{Rado}.
Hence, its $V$-representation reads

\begin{align}
 \mathcal{M}^{LU}_s(\psi) = \text{conv}\left ( \{\lambda(\psi)_{\sigma}\}_{\sigma \in \Sigma_d}\right). \label{eq:Rado}
\end{align}

Note further that, as the cardinality of $\Sigma_d$ is $d!$, every generic state in $\mathcal{M}_s(\psi)$, i.e. every state
with pairwise different Schmidt components, is represented $d!$ times in
$\mathcal{M}^{LU}_s(\psi)$. This does not hold for non-generic states. However, as the set of non-generic states is of measure zero,
we obtain that
\begin{align}
 V_s(\psi) = \mu(\mathcal{M}_s(\psi)) = \frac{1}{d!}\mu(\mathcal{M}^{LU}_s(\psi)).
\end{align}
We are going to show below that $\mathcal{M}^{LU}_s(\psi)$ is generically a $(d-1)$-dimensional polytope. Hence, we choose for $\mu$ the Lebesgue measure
in $\R^{d-1}$ in order to arrive at a physically meaningful quantification of the source set. The non-generic case will be treated afterwards (see Appendix A).

In order to determine now the volume of $\mathcal{M}^{LU}_s$ we use the results presented in \cite{Brion}.
There, the volume of a simple polytope, $P \subset \R^k$, with $\text{dim}(P) = k$, was shown to be

\begin{align}
\label{Eq_Volume}
\mu(P)=\frac{(-1)^k}{k!} \sum_{v \in \text{vert}(P)} \abs{\text{det}(m_v)}\frac{\langle v, \xi \rangle^k}{\prod\limits_{i=1}^k  \langle e_i(v), \xi \rangle}.
\end{align}

Here, $m_v$ is the $k\times k$ matrix $[e_1(v); e_2(v); \ldots; e_k(v)]$, where $e_i(v) = v - v_i(v)$ denotes the edge vector that connects
the vertex $v$ with its $i$-th neighboring vertex $v_i(v)$ \footnote{Note that the order of the neighbors is not important as the absolute value of the
determinant is considered. Note further that
from the expression for $\mu(P)$ in Eq.(\ref{Eq_Volume}) it is evident that this formula can only be used to determine the volume of a simple
polytope, as only in case each vertex has exactly $k$ neighbors all the matrices $m_v$ occurring in Eq. (\ref{Eq_Volume}) are square matrices.}.
Moreover, the vector $\xi \in \R^k$ is an arbitrary vector with the property that it is not orthogonal to any edge vector, i.e.
$\langle \xi, e_i(v)\rangle \neq 0$ for all $v$ and $i$. In fact, $\abs{\text{det}(m_v)}/k!$ is the volume
of the simplex defined by the vertex $v$ and its neighbors. With the term simplex we refer to a $k$-dimensional polytope in $\R^k$ that
is the convex hull of $k+1$ vertices. The volume of the whole polytope $P \subset \R^k$ is thus the weighted sum of these different, but overlapping, simplices.

Hence, in order to compute $V_s(\psi)$ via Eq. (\ref{Eq_Volume}) it remains to show that $\mathcal{M}^{LU}_s(\psi)$ is a simple polytope and to find the vector $\xi$ with
the properties mentioned above. Note that, due to the normalization condition, $\sum_{i=1}^{d} \lambda_i = 1$, we have that $\mathcal{M}^{LU}_s(\psi)$ is at
most a $(d-1)$-dimensional polytope in $\R^d$. As a consequence, we need to identify the $(d-1)$-dimensional subspace of $\R^d$, ${\cal A}$, which contains
$\mathcal{M}^{LU}_s(\psi)$ and show that $\mathcal{M}^{LU}_s(\psi)$ is simple. More specifically, we are going to show
that every vertex of $\mathcal{M}^{LU}_s(\psi)$ has generically $(d-1)$ neighbors and that the vertex and its neighbors are affinely independent, i.e. the $(d-1)$
edge vectors $\{e_i(v)\}_i$ are linearly independent. This implies that $\mathcal{M}^{LU}_s(\psi)$ is a $(d-1)$-dimensional simple polytope. Finally,
we have to find $\xi$ fulfilling the properties mentioned above.

Let us first describe the $(d-1)$-dimensional subspace of $\R^d$, which contains $\mathcal{M}^{LU}_s(\psi)$.
Note that $\mathcal{M}^{LU}_s(\psi) \subset \{\lambda\in \R^d \ \mbox{s.t.} \ \sum_i \lambda_i=1\}=:{\cal A}$, where ${\cal A}=\frac{1}{\sqrt{d}}\hat{\phi}^+ +U$, is a
$(d-1)$-dimensional
affine subspace of $\R^d$ \footnote{In the following we use a transformation into the subspace $\cal{A}$ in order to calculate the volume of
the polytope. Another approach would be to look at the polytope in $\R^{d-1}$, defined by vectors $\lambda = (\lambda_1,\ldots\lambda_{d-1})$ that
fulfill the normalization condition $\sum_{i=1}^{d-1}\lambda_i \leq 1$. The resulting polytope would be a projection of the
initial one onto the space spanned by the first $(d-1)$ standard basis vectors. Hence, its volume would be only $1/\sqrt{d}$-times as big as the original
volume (see Appendix B). In fact, in Sec. \ref{sec:Accessible} we use this approach to calculate the accessible volume. However, in the case
of the source volume this approach is more involved than the one presented here.}. Here, $\hat{\phi}^+=\frac{1}{\sqrt{d}}(1,1,\ldots, 1)$ and

\begin{align} U=\{\lambda \in \R^d \ \mbox{s.t.} \  \sum_i \lambda_i=0\}.\label{eq:DefU}\end{align}

We consider first the generic case. That is, there exists
no pair of different indices, $i,j$, such that $\lambda_i=\lambda_j$. Notice then that in this case the vertices of this polytope are in 
one-to-one correspondence with those of the so-called permutahedron, i.e. the polytope whose vertices are formed by all permutations of the 
coordinates of the vector $(1,2, \ldots,d)$, which has been well-studied in polytope theory. It is known that the 
permutahedron is simple and its vertices have been characterized \cite{Ziegler}. For the sake of readability, we give 
here a self-contained proof that $\mathcal{M}^{LU}_s(\psi)$ corresponds generically to a simple polytope in the $(d-1)$-dimensional subspace 
${\cal A}$, i.e. that every vertex has exactly
$d-1$ affine independent neighbors.\\
First of all, we determine the
neighbors of the vertex $\lambda(\psi) = \lambda^\downarrow(\psi)$, which we assume to be sorted already. To do so, we
consider here the set of
$d-1$ independent inequalities $\varOmega_1=\{E_k(\lambda')\leq E_k(\lambda(\psi)) \ \mbox{s.t.} \ k\in \{1,\ldots, d-1\}\}$ and the set
$\varOmega_2= \varOmega\backslash \varOmega_1$, where $\varOmega$ is as defined in Eq. (\ref{Ineq_Majunsort}). Clearly, $\lambda(\psi)$ trivially satisfies all $d-1$ inequalities in
$\varOmega_1$ as equalities. Its neighboring vertices must satisfy $d-2$ inequalities in $\varOmega_1$ and one independent inequality of
$\varOmega_2$ as equalities, as they must be elements of $d-2$ facets of which $\lambda(\psi)$ is an element too. Using now that each of the
vertices must be of the form
$\lambda(\psi)_{\sigma}$ (see Eq. (\ref{eq:Rado})), we have that the neighbors of $\lambda(\psi)$ can only be of the form
$\lambda_{\mathbb{1},i}=P_{\tau_{i,i+1}}\lambda(\psi)$, where $P_{\tau_{i,i+1}}$
denotes the permutation matrix permuting component $i$ with component $i+1$ and leaving the rest unchanged, e.g.
$P_{\tau_{1,2}}\lambda(\psi) =(\lambda_2,\lambda_1,\lambda_3,\ldots, \lambda_d)$. As there are $d-1$ possible permutations of this kind, we have $d-1$
neighbors of the vertex $\lambda(\psi)$ \footnote{Note that the vector $P_{\tau_{1,d}}\lambda(\psi) = (\lambda_d,\lambda_2,\ldots,\lambda_{d-1},\lambda_1)$
is not a neighbor of $\lambda(\psi)$ as
it does not fulfill any of the inequalities in $\varOmega_1$ as equality.}. As the $d-1$ edge vectors $e(\lambda(\psi))_i = \lambda(\psi) - \lambda(\psi)_i$
are of the form
$e(\lambda(\psi))_i = (0,\ldots,0,\lambda_{i+1}-\lambda_i,\lambda_i - \lambda_{i+1},0,\ldots,0)$, they
span a $(d-1)$ dimensional space. Hence the polytope $\mathcal{M}^{LU}_s(\psi)$ is indeed $(d-1)$-dimensional. The same argument holds for any other vertex $\lambda(\psi)_\sigma$, which has
the neighbors $\lambda_{\sigma,i}=P_{\sigma}P_{\tau_{i,i+1}}\lambda(\psi)$ \footnote{This can be easily seen as follows. We take an arbitrary vertex
$\lambda(\psi)_{\sigma}$ of the source set.
This vertex fulfills the inequalities $E^{\sigma^{-1}}_k(\lambda(\psi)_{\sigma}) \leq E_k(\lambda(\psi)), k \in \{1,\ldots,d-1\},$
as an equality. Based on our previous results we therefore conclude that the
vertices of the form $P_{\sigma}P_{\tau_{i,i+1}}P_{\sigma^{-1}}P_{\sigma}\lambda(\psi) = P_{\sigma}P_{\tau_{i,i+1}}\lambda(\psi) = \lambda(\psi)_{\sigma,i}$ are the only neighboring
vertices of $\lambda(\psi)_\sigma$ for $k \in \{1,\ldots,d-1\}$.}.
Hence, all vertices have exactly $d-1$ neighbors, which implies that the $(d-1)$-dimensional polytope is simple. Note that it does not need to be simple in case of
degeneracy, i.e. for non-generic states. However, using continuity arguments, one can show that the volume of the corresponding polytope can be computed
as in case of no degeneracy (see Appendix A for details).\\

We can now use the volume formula given in Eq. (\ref{Eq_Volume}) in order to calculate the source volume of a generic state with Schmidt vector
$\lambda(\psi) = \lambda^{\downarrow}(\psi)$. The volume is translationally invariant, implying that
$V^{LU}_s(\psi) = \mu(\mathcal{M}^{LU}_s(\psi)) = \mu(\mathcal{M}^{LU}_s(\psi) - 1/\sqrt{d} \hat{\phi}^+)$. The vertices
of the translated polytope are given by $\lambda(\psi)_{\sigma} - 1/\sqrt{d} \hat{\phi}^+$.
Note that these vertices are $d$-dimensional vectors. However, they obviously all belong to the $(d-1)$-dimensional subspace $U$ defined in
Eq. (\ref{eq:DefU}). In order to use Eq. (\ref{Eq_Volume}) all vectors needed there must be considered as linear combinations of an orthonormal
basis of $U$, e.g.
the matrix $m_{v}$ contains the coordinates of the edge vectors $\{e_i(v)\}_i$ with respect to that basis. However, we use
the following properties to circumvent this basis change. Note that the term $\abs{\text{det}(m_v)}$
in Eq. (\ref{Eq_Volume}) gives the volume of the $k$-dimensional parallelotope that is defined by the edge vectors $\{e_i(v)\}_i$. Its numerical
value is identical to the volume of a $(k+1)$-dimensional parallelotope of which the $k$ vectors that define the base coincide with the original ones and
the $(k+1)$-th vector is normalized and orthogonal to the subspace that contains the base, e.g. the area of a rectangle is equal to the volume of a
cuboid that is of height one and whose base coincides with the rectangle \footnote{Indeed, it holds for any matrix
$A \in \R^{k \times k}$ that $\abs{\text{det}(A)} = \abs{\text{det}( 1 \oplus A)}$, where the left-hand-side is the volume of a $k$-dimensional
parallelotope
defined by the column-vectors of $A$, and the right-hand-side is the volume of a $(k+1)$-dimensional polytope that has the previous
polytope as base to which the first vertex is orthogonal.}. For the current computation this means that
\begin{align}
\abs{\text{det}(m_{\lambda_{\sigma}})} = \abs{\text{det}\left(\hat{\phi}^+;e_1(\lambda_{\sigma});e_2(\lambda_{\sigma});\ldots;e_{d-1}(\lambda_{\sigma})\right)}, \nonumber
\end{align}
where $\hat{\phi}^+$, as defined before, is the vector that is orthogonal to the edge vectors
$e_i(\lambda_{\sigma}) = P_\sigma e(\lambda(\psi))_i$, with
$e(\lambda(\psi))_i= (0,\ldots,0,\lambda_{i+1}-\lambda_i, \lambda_i-\lambda_{i+1},0,\ldots,0)$. That is, we do not need to express the
edge vectors in terms of an orthonormal basis of $U$ in order to compute this determinant, but we can continue with the representation in
terms of their $d$ coordinates in the standard basis of $\R^d$. Using the
fact that $P_\sigma\hat{\phi}^+ = \hat{\phi}^+$ and $\abs{\text{det}(P_{\sigma})} = 1$ it follows that
\begin{align}
&\abs{\text{det}(m_{\lambda_{\sigma}})} = \nonumber \\
&=\abs{\text{det}(P_{\sigma})} \abs{\text{det}\left(\phi^+;e_1(\lambda(\psi));e_2(\lambda(\psi));\ldots;e_{d-1}(\lambda(\psi))\right)} \nonumber\\
&= \sqrt{d} \prod\limits_{k=1}^{d-1} \abs{\lambda_k - \lambda_{k+1}}.
\end{align}
The last expression can be easily obtained as each vector $e_i(\lambda(\psi))$ contains only two non-vanishing entries.
It is now easy to choose a vector $\xi \in U$ fulfilling that it is non-orthogonal to any of the edge vectors $e_i(v)$, as required by Eq. (\ref{Eq_Volume}).
Note that $\xi$ only appears in inner products. Moreover, the inner product is invariant under a basis change from an orthonormal basis of $U$,
extended by $\phi_+$ in order to be a basis of $R^d$, and the standard basis of $\R^d$.
Consequently, we can again represent $\xi$ and all other vectors in these inner products in the standard basis. In fact, for $\xi = (1,2,\ldots,d) - \frac{d+1}{2}(1,1,\ldots,1) \in U$ it holds that
\begin{align}
\langle \xi, e_i(\lambda_{\sigma})\rangle &= \langle \xi, P_{\sigma}e_i(\lambda(\psi))\rangle = \langle P_{\sigma^{-1}}\xi, e_i(\lambda(\psi))\rangle \nonumber\\
&= \left(\sigma^{-1}(i) - \sigma^{-1}(i+1)\right)\left(\lambda_{i+1} - \lambda_{i}\right),
\end{align}
where we used that $P_{\sigma}^T = P_{\sigma^{-1}}$.
Furthermore, we have $\langle \xi, \lambda(\psi)_{\sigma} - 1/\sqrt{d} \hat{\phi}^+ \rangle = \sum_{k=1}^d \sigma^{-1}(k) \lambda_{k} - \frac{d+1}{2}$.
Plugging these expressions into the volume formula in Eq. (\ref{Eq_Volume}), relabeling $\sigma^{-1}$ with $\sigma$, using that
$V_s(\psi) = \frac{1}{d!}\mu(\mathcal{M}^{LU}_s(\psi))$ and the fact that $\lambda(\psi)$ is a sorted Schmidt vector, we obtain
\begin{align}
 V_s(\psi)  = \frac{1}{d!}\frac{\sqrt{d}}{(d-1)!} \sum\limits_{\sigma \in \Sigma_d} \frac{\left(
\sum\limits_{k=1}^d \sigma(k) \lambda_k - \frac{d+1}{2}\right)^{d-1}}{\prod\limits_{k=1}^{d-1} \sigma(k) - \sigma(k+1)}. \label{eq:SourceVolume}
\end{align}
We show in Appendix A that this formula holds also in the case of degenerate Schmidt coefficients.
Note that the separable state, $\ket{\psi}_{sep}$, with Schmidt vector $(1,0,0,\ldots,0)$ can be obtained from any other quantum state via LOCC and therefore 
maximizes the source volume, i.e. $\sup_{\phi \in \mathcal{H}}  V_s(\phi) = V_s(\psi_{sep})$. According to Eq. (\ref{eq:Rado}) the vertices of the source set of $\ket{\psi}_{sep}$ are given by the standard vectors 
$e_i$ in $\R^d$. $\mathcal{M}_s^{LU}(\psi_{sep})$ is thus a simplex defined by the $d$ vertices $\{e_i\}_{i=1}^d$ whose volume is easily computed. It is then straightforward to see that $V_s(\psi_{sep}) = \sqrt{d}/(d! (d-1)!)$. Furthermore, it is easily 
seen that $V_s(\phi^+) = 0$, as we show in Appendix A. Using the definition of the source entanglement, i.e.
$E_s(\psi) = 1 - \frac{V_s(\psi)}{\sup\limits_{\phi \in \mathcal{H}}  V_s(\phi)}$,
we obtain the following lemma.
\begin{lemma}
\label{lem:SourceEnt}
The source entanglement of a bipartite state, $\ket{\psi}\in \C^d \otimes \C^d$ with sorted Schmidt vector $\lambda(\psi)$ is given by
\begin{align}
 E_s(\psi) = 1- \sum\limits_{\sigma \in \Sigma_d} \frac{\left(
\sum\limits_{k=1}^d \sigma(k) \lambda_k - \frac{d+1}{2}\right)^{d-1}}{\prod\limits_{k=1}^{d-1} \sigma(k) - \sigma(k+1)}.
\end{align}
\end{lemma}

In what follows we show how one can generalize the formula for the source entanglement in Lemma \ref{lem:SourceEnt} in order to obtain a whole class 
of new operational entanglement measures as mentioned in Section \ref{sec:AccandSourc}.\\

\paragraph*{Generalizations of the source entanglement} \hfill\\
\hfill\\
We consider, for given $\ket{\psi}$
with sorted Schmidt vector $\lambda(\psi)$, its source set of states $\ket{\Psi} \in \C^k \otimes \C^k$, where $k \geq d$. More specifically, we measure the set
of states with greater or equal local dimensions that can be converted to $\ket{\psi}$ via LOCC. This set is denoted by $M_s^k(\psi)$ and reads
\begin{align}
M_s^k(\psi) = \{\ket{\Psi} \in \C^k \otimes \C^k \ \mbox{s.t.} \ \ket{\Psi} \xrightarrow{LOCC} \ket{\psi}\}.
\end{align}
In fact, we can identify $\ket{\psi}$ with a state $\ket{\Psi^k(\psi)} \in \C^k \otimes \C^k$ that has the Schmidt vector
$\lambda(\Psi^k(\psi)) = (\lambda_1,\ldots,\lambda_d,0,\ldots,0) \in \R^k$, where we simply appended $k-d$ zeros to the initial, $d$-dimensional
Schmidt vector of $\ket{\psi}$. The volume corresponding to $M_s^k(\psi)$ is then given by
$V_s^k(\psi) = V_s(\Psi^k(\psi))$. In this way we obtain a whole class of operational
entanglement measures that are a generalization of the source entanglement presented in Lemma \ref{lem:SourceEnt} and that read
\begin{align}
E_s^k(\psi) = \frac{1}{\sup\limits_{\ket{\phi} \in \mathcal{H}} E_s(\Psi^k(\phi))} E_s(\Psi^k(\psi)) \quad k \geq d. \label{eq:SourceGeneralization}
\end{align}
Here, we divided by the supremum over all $\ket{\phi} \in \C^d \otimes \C^d$ such that the range of $E_s^k(\psi)$ is $[0,1]$. 
Note that these generalizations include the initial source entanglement for $k=d$. $E_s^k(\psi)$ is a polynomial of degree at most $k-1$ in $d-1$ Schmidt
coefficients of $\ket{\psi}$, where the remaining coefficient is given by normalization. 
We thus have explicit formulae for a whole class of operational entanglement measures that quantify the set of LU-inequivalent, bipartite pure states with the 
same or higher dimension that can be transformed to a given bipartite  
state,
$\ket{\psi}$, via LOCC. Stated differently, the set of measures
 \begin{align}
 \{E_s^k(\psi)\}_{k \geq d}.
 \end{align}
characterizes how easy it is to generate a single copy of $\ket{\psi}$ from a single copy of another bipartite pure state via LOCC.

\subsection{Accessible volume}
\label{sec:Accessible}

We consider now the accessible set given in Eq. (\ref{Eq_MsMa1}) for a state $\ket{\psi}$ whose sorted Schmidt vector is given by $\lambda(\psi) =
\lambda^\downarrow(\psi)$.
As seen from Eq. (\ref{Eq_MsMa1}) this set is the intersection of the halfspaces defined by the inequalities
\begin{align}
\label{eq:Ma}
&E_k(\lambda')\geq E_k(\lambda(\psi)) \quad \forall k\in \{1,\ldots, d-1\}, \\
&\lambda_1'\geq \lambda_2' \geq \ldots \geq \lambda_d'\geq 0.
\end{align}

Our aim is now to compute the volume of this convex polytope. Let us first note that we consider here only the LU-equivalence classes as we are fixing the order of the Schmidt coefficients.
Any other fixed order would obviously also represent all LU classes, despite the fact that the corresponding vectors belong to a different
subset of $\R^d$. Moreover, the volume of these different sets is the same. In contrast to the source set, they are, however, not connected in general and
the union of these sets of differently sorted vectors that majorize
$\lambda(\psi)$ would not lead to a convex set. In order to see that consider $\lambda(\psi)$ and $\lambda' = (1,0,\ldots,0)$, i.e. the Schmidt vector
corresponding to a separable state. Clearly, it holds that $\lambda' \succ \lambda(\psi)$. However, the set is in general not convex, as $1/2 \lambda' + 1/2 P_{\tau_{1,2}} \lambda' \nsucc \lambda$ if $\lambda_1 > 1/2$. The above mentioned properties of the union of the different representations of the accessible set is the reason why we do not consider this set,
although we used such an approach in case of the source set.
However, the accessible set defined by the halfspaces in Eq. (\ref{eq:Ma}) is, as the source set was, a convex set.

Let us now demonstrate how the volume of $\mathcal{M}_a(\psi)$ can be computed. To do so, we first determine the vertices of the accessible set.
Thereby, it is useful to get rid of the last Schmidt coefficient,
which is determined by the normalization condition, i.e. $\lambda_d = 1 - E_{d-1}(\lambda)$. Hence, from now on, we consider the Schmidt vectors as $d-1$ dimensional vectors and the
$H$-representation of the accessible set is given by
\begin{align}
 &E_k(\lambda')\geq E_k(\lambda(\psi)) \quad \forall k\in \{1,\ldots, d-1\}, \label{eq:H-RepAcc}\\
 &\lambda_1' \geq \ldots \geq \lambda_{d-1}' \geq 1 - E_{d-1}(\lambda') \geq 0. \label{eq:Ordering}
\end{align}
A vertex of $M_a$, $v_i$, is the unique vector in $M_a \subset \R^{d-1}$, where $d-1$ independent inequalities are satisfied as equations.
Let us remark here that the method to calculate the vertices of the accessible set and its volume differs from the one used to calculate the source volume, as we will explain in the following. To determine the source volume, we calculated the $(d-1)$-dimensional volume of the source set that is contained in $A \subset \R^d$ (respectively $U$, after a translation). This is why we had to, at least conceptually, transform into an orthonormal basis of $U$. To calculate the accessible volume, however, we abandoned the last
Schmidt coefficient. As a result, the $H$-representation in Eqs. (\ref{eq:H-RepAcc}) and (\ref{eq:Ordering}) describes the projection of the accessible volume that is contained
in $A \subset \R^d$ onto the subspace spanned by the first
$(d-1)$ standard basis vectors. Subsequently we identified this subspace with $\R^{d-1}$. As shown in Appendix B, the resulting volume is $1/\sqrt{d}$-times 
as big as the original one. However, this constant factor is at the end irrelevant as we rescale the volume in order to obtain the accessible 
entanglement (see Eq. (\ref{eas})). 

Let us now consider the determination of the vertices of the accessible set. Note that $\lambda(\psi)$ is always a vertex as it obeys all $d-1$ inequalities in Eq. (\ref{eq:H-RepAcc}) as equalities.
As we have $2d-1$ inequalities, there can exist at most $\binom{2d-1}{d-1}$ vertices.
Clearly, there will be much less in general.
Consider for instance for $d=4$ the state $1/100 \cdot (30,27,24,19)$ that has ten vertices, while the state $1/10 \cdot (4,3,2,1)$ has only eight, which is
a lot less than $\binom{7}{3} = 35$ . As can be seen from these examples, even the number of vertices
depends also for non-degenerate states very strongly on $\lambda(\psi)$. Furthermore, it is not clear if the accessible set is always a simple 
polytope for non-degenerate states. These are probably the reasons why it is difficult to derive a closed expression for the
vertices or the volume of the accessible set given some state corresponding to the Schmidt vector $\lambda(\psi)$. 
The vertices can however be easily computed for a given $\lambda(\psi)$, as we will briefly explain in the following.

Assume that one vertex, $v_1$, is known (e.g. $\lambda(\psi)$ in the case of the accessible set). Let $\tilde{\varOmega}_{1}$ denote a set of
$d-1$ independent inequalities which are fulfilled as equalities for the given vertex. In order to find a neighboring vertex, pick one
inequality which is in $\tilde{\varOmega}_{1}$ and replace it by an inequality from the $H$-representation which is not in $\tilde{\varOmega}_{1}$.
The vector in $\R^{d-1}$ satisfying these inequalities now as equalities (and of course obeying all other inequalities) is a neighbor of $v_1$,
as it is an element of $d-2$ facets that also contain $v_1$. One can continue in this way until one finds all possible vertices.

In the literature there exist several algorithms for computing the vertices and the volume of arbitrary convex sets. The reader is referred to
\cite{Bueler} for a review on the basic properties of some algorithms for volume computation.
Usually the volume is computed by using either the so-called triangulation method or the signed decomposition method. In the former method the polytope
is decomposed into simplices with mutually disjoint interior such that the volume of the whole polytope is the sum of the volumes of the individual
simplices.
In the latter method the simplex is decomposed into signed simplices, i.e. simplices to which a positive or negative sign is associated, that are allowed to overlap. These are then, depending on their sign, added or
subtracted successively in order to obtain the volume of the polytope. We will use here the algorithm presented in \cite{avis} to
calculate all vertices of the accessible polytope. Although one can find the vertices using the method outlined above, the aim is of course to
perform the computation using few resources.
In \cite{avis} it has been shown that, by choosing the inequalities that have to be interchanged as explained above according to
certain rules, the vertices can be found easily. In fact the algorithm finds the $N$ vertices of a polytope in $\R^{d-1}$ defined by a nondegenerate system 
of $2d-1$ inequalities in time $\mathcal{O}(d^2N)$ and $\mathcal{O}(d^2)$ space. Note, however, that $N$ can in principle grow exponentially with 
$d$. A revised version of the algorithm presented in \cite{avis2} is also capable of
computing the volume of the convex polytope using the triangulation method described above.

Whereas the numerical methods described above can be utilized to compute the accessible volume for arbitrary states, we present in Sec. \ref{sec:Examples} some examples of bipartite systems of small dimension for which we computed analytic
expressions for the vertices, the corresponding accessible
volume and the associated accessible entanglement. Note that, similarly to the source entanglement, also the accessible entanglement can be generalized by 
considering different Hilbert spaces of the initial state and the state in the accessible set (\cite{us}, see Sec. \ref{sec:Examples} for examples).

Before concluding this subsection, let us present some states which always correspond to vertices of the accessible polytope. 
The reason for doing so is that the knowledge of the vertices of the accessible set allows us to gain more insight into the entanglement properties of the state. The following vectors are always vertices of the accessible set of a state with sorted
Schmidt vector $\lambda(\psi) = (\lambda_1,\ldots,\lambda_d)$,
\begin{align}
 &v_1 = (\lambda_1,\lambda_1, \ldots, \lambda_1, \lambda_{norm,1},0,\ldots,0),\\
 &v_2 = (\lambda_1,\lambda_2, \lambda_2, \ldots, \lambda_2, \lambda_{norm,2},0, \ldots,0),\\
 &\ldots \nonumber\\
 &v_{d-2} = (\lambda_1, \lambda_2, \lambda_3,\ldots,\lambda_{d-2},\lambda_{d-2}, \lambda_{norm,d-2}),
\end{align}
where $0 \leq \lambda_{norm,i} \leq \lambda_i$ is such that all components add up to one. The reason for that is that
$v_i$ obviously majorizes $\lambda(\psi)$ and is an element of $d-1$ facets of the accessible set. More precisely, $v_i$ 
satisfies $i$ of the inequalities from Eq. (\ref{eq:H-RepAcc}) and $d-1-i$ inequalities from
Eq. (\ref{eq:Ordering}) as equalities. Note that these $d-1$ inequalities are independent such that $v_i$ is indeed a vertex. Furthermore, all maximally entangled states of two $k$-level systems, $\ket{\phi^+_k}$ , i.e.
states where the first $k$ entries of the sorted Schmidt vector are $1/k$ and the remaining ones are zero, are in the accessible set iff
$\lambda_{1}\leq 1/k$. To see this, note that this criterion is equivalent to $E_1(\phi^+_k) \geq E_1(\lambda(\psi))$. Since $\lambda(\psi)$ is sorted, this
implies that $E_i(\phi^+_k) \geq E_i(\lambda(\psi))$ for all $i$ and therefore (see Eq. (\ref{Ineq_Maj})) that $\ket{\phi^+_k}$ is accessible
from $\ket{\psi}$. It is straightforward to show that $\lambda(\phi^+_k)$ is in this case also a vertex of the accessible polytope that fulfills $d-1$ of
the inequalities in Eq. (\ref{eq:Ordering}) as equalities.

\subsection{Source and accessible volume for low dimensional bipartite systems}
\label{sec:Examples}

In this section we use the results obtained above in order to compute the accessible as well as the source entanglement of bipartite systems.
More specifically, we present the source and accessible set of small dimensional systems, i.e. $d=2,3,4$ in Figs. \ref{fig:2x2sets} - \ref{fig:4x4sets} 
and briefly discuss their properties. By means of these figures we can also derive closed expressions for the accessible volume of a state of two qubits or
two qutrits.

\subsubsection{Two qubits}

\begin{figure}[h!]
 \includegraphics[width=0.3\textwidth]{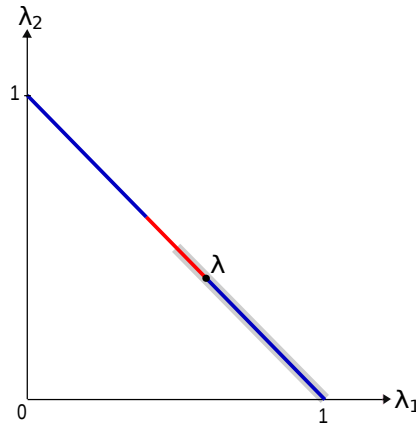}
 \caption{(colors online) The source and accessible set of a two-qubit state with sorted Schmidt vector $\lambda=(0.6,0.4)$. The line that connects the
 points $e_1=(1,0)$ and $e_2=(0,1)$ corresponds to the
 set of valid, i.e. normalized, Schmidt vectors. The shaded region highlights
 the set of sorted Schmidt vectors that is in one-to-one correspondence to the LU-equivalence classes. The red (blue) lines in this region depict the source
 (accessible) volume of the quantum state respectively.}
 \label{fig:2x2sets}
\end{figure}

In the simplest example of a two-qubit state there exists only one independent Schmidt coefficient, such that the source and accessible volume of
a state with sorted Schmidt vector $\lambda(\psi) = (\lambda, 1 - \lambda)$ read
\begin{align*}
&\mathcal{M}_s(\psi) = \{ (\lambda'_1,1-\lambda'_1) \ \mbox{s.t.} \ 1/2 \leq \lambda'_1 \leq \lambda\},\\
&\mathcal{M}_a(\psi) = \{ (\lambda'_1,1-\lambda'_1)\ \mbox{s.t.} \ \lambda \leq \lambda'_1 \leq 1\}.
\end{align*}
Consequently, the two sets are complementary in the sense that their union results in the whole state space and there do not exist LOCC-incomparable states. That is, given two states one
of them can always be converted into the other via LOCC. For this simple example the volumes can be directly read off from Figure \ref{fig:2x2sets}, in which
both sets are depicted for a given state. It is easy to see that the source and accessible volume of a two-qubit state $\ket{\psi}$ are given by
\begin{align*}
&V_s(\psi) = \sqrt{2}\left(\lambda_1 - 1/2\right),\\
&V_a(\psi) = \sqrt{2}\left(1-\lambda_1\right).\\
\end{align*}
Using these formulas it is straightforward to see that the source and accessible entanglement (see Eqs. (\ref{eas})) of a two-qubit state are in fact identical and given
by
\begin{align}
E_s(\psi) = E_a(\psi) = 2(1-\lambda_1). \nonumber
\end{align}
Clearly, each of them gives a unique characterization of the entanglement contained in a two-qubit state, i.e. they allow to recover its Schmidt coefficients. \\

\subsubsection{Two qutrits}

\begin{figure}[h!]
 \includegraphics[width=0.3\textwidth]{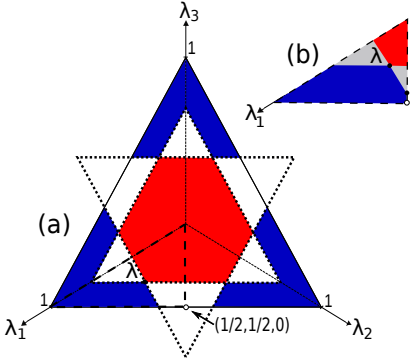}
 \caption{(colors online) (a) The source and accessible set of a two-qutrit state with sorted Schmidt vector $\lambda = (0.6,0.37,0.13)$.
 The triangle with the three vertices $e_1 = (1,0,0)$, $e_2 = (0,1,0)$ and $e_3 = (0,0,1)$ corresponds to all valid, i.e. normalized, Schmidt vectors.
 The thick, dashed line encloses the set of sorted Schmidt vectors that is in one-to-one correspondence to the LU-equivalence classes. The red (blue) regions depict the source (accessible)
 set of the quantum state respectively.  The hyperplanes that correspond to the halfspaces
 defining the set $\mathcal{M}^{LU}_s(\psi)$ are also indicated by the dotted lines. The white dot corresponds to the maximally entangled state $\ket{\phi^+_2}$ 
 of two qubits with Schmidt vector $(1/2,1/2,0)$. Since the depicted state has $\lambda_1 > 1/2$ the state $\ket{\phi^+_2}$ is not an element of 
 its accessible set. Its accessible set has four vertices.\\
 (b) The source and accessible volume of a two-qutrit state with sorted 
 Schmidt vector $\lambda = (0.47,0.36,0.17)$. In contrast to the state in (a), this state fulfills $\lambda_1 < 1/2$ s.t. the source volume contains the state $\ket{\phi^+_2}$ and 
 the accessible volume has five vertices.}
  \label{fig:3x3sets}
\end{figure}

Let us proceed with the investigation of a system of two qutrits. In contrast to the two-qubit case one can find two-qutrit states that are LOCC-incomparable. Consequently, the
source and accessible set of a generic two-qutrit state are not complementary, as can be seen in Fig. \ref{fig:3x3sets}. More precisely, their union
does not lead to the whole state space. Moreover one can easily see that representations of
the accessible set that correspond to different orderings of the Schmidt coefficients are not connected to each other. In contrast to that
the union of the corresponding representations of the source set, i.e. $\mathcal{M}^{LU}_s(\psi)$, is a convex set. We use Eq. (\ref{eq:SourceVolume}) 
to obtain the source volume of a state $\ket{\psi}$. Based on Fig. \ref{fig:3x3sets} we can also 
calculate its accessible volume using elementary geometry. That is, we subdivide the accessible set into mutually disjoint triangles. The shape 
of the accessible set, and with it the triangulation, depends on the largest Schmidt vector, $\lambda_1$, of the state. This can 
be easily seen in Fig. \ref{fig:3x3sets}. The 
volume of a triangle with vertices $v_1, v_2$ and $v_3$ is then given by the formula $1/2 \abs{(v_2-v_1) \times (v_3-v_1)}$, where $a \times b$ denotes the 
cross product between vectors $a,b \in \R^3$. 
In analogy to the generalization of the source volume
discussed in Sec. \ref{sec:SourceVolume} we moreover
compute $V_a^2(\psi)$, i.e. the volume of accessible states with at least one zero Schmidt coefficient. Stated differently, we compute the volume
of two-qubit states that is accessible from $\ket{\psi}$.
We obtain the volumes
\begin{align*}
&V_s(\psi) = \frac{6 \lambda_2 \lambda_3-3 \lambda_2^2+6 (\lambda_3-1) \lambda_3+1}{4 \sqrt{3}},\\
&V_a(\psi) =   \begin{dcases}
    \sqrt{3}\lambda_2\lambda_3 & \textrm{if } \lambda_1 > \frac{1}{2} \\
    \sqrt{3} \left[ \lambda_2\lambda_3-\frac{1}{4}\left( 1-2\lambda_1
    \right)^2 \right] & \textrm{if } \lambda_1\leq\frac{1}{2}
  \end{dcases}\\
&V_a^2(\psi) =  \begin{dcases}
    \sqrt{2}(1-\lambda_1) & \textrm{if } \lambda_1 > \frac{1}{2} \\
    \frac{\sqrt{2}}{2}  & \textrm{if } \lambda_1\leq\frac{1}{2}.
  \end{dcases}
\end{align*}
Starting from a state $\ket{\psi}$ with three nonzero Schmidt coefficients one can, as pointed out in Sec. \ref{sec:Accessible}, access the maximally
entangled state of two-qubits, $\ket{\phi^+_2}$, with Schmidt vector $\lambda(\phi^+_2)=(1/2,1/2,0)$, iff $\lambda_1 \leq 1/2$. It is clear that in this
case one can obtain any two-qubit state. This is why such a state maximizes $V_a^2(\psi)$.
It is straightforward to compute the entanglement measures from the volumes above. We obtain
\begin{align*}
&E_s(\psi) = 3\lambda_2^2-6\lambda_2 \lambda_3 - 6 (\lambda_3-1)\lambda_3,\\
&E_a(\psi) =   \begin{dcases} 12 \lambda_2 \lambda_3 & \textrm{if } \lambda_1 > \frac{1}{2} \\
12 [\lambda_2 \lambda_3 - 1/4 (1-2\lambda_1)^2] & \textrm{if } \lambda_1\leq\frac{1}{2}
  \end{dcases}\\
&E_a^2(\psi) =  \begin{dcases}
    2(1-\lambda_1) & \textrm{if } \lambda_1 > \frac{1}{2} \\
    1  & \textrm{if } \lambda_1\leq\frac{1}{2},
  \end{dcases}
\end{align*}
 where we defined $E_a^2(\psi) = \frac{V_a^2(\psi)}{\sup_\phi V_a^2(\phi)}$. Using Eq. (\ref{eq:SourceGeneralization}) one can furthermore calculate
\begin{align}
E^4_s(\psi) = \frac{27}{13} \left(2 \lambda_2^3+6 \lambda_2^2 \lambda_3+3 (3-4 \lambda_2) \lambda_3^2-10 \lambda_3^3\right).\nonumber
\end{align}
Note that one can show that $E_s(\psi)$ together with $E^4_s(\psi)$ uniquely characterize the entanglement contained in a 
quantum state of two qutrits. That is, these two measures uniquely define the Schmidt components of the state. Hence, its entanglement is completely 
characterized by how many two-qutrit states and states of two four-level systems can reach the state at hand.

\subsubsection{Two four-level systems}

\begin{figure}[h!]
 \includegraphics[width=0.3\textwidth]{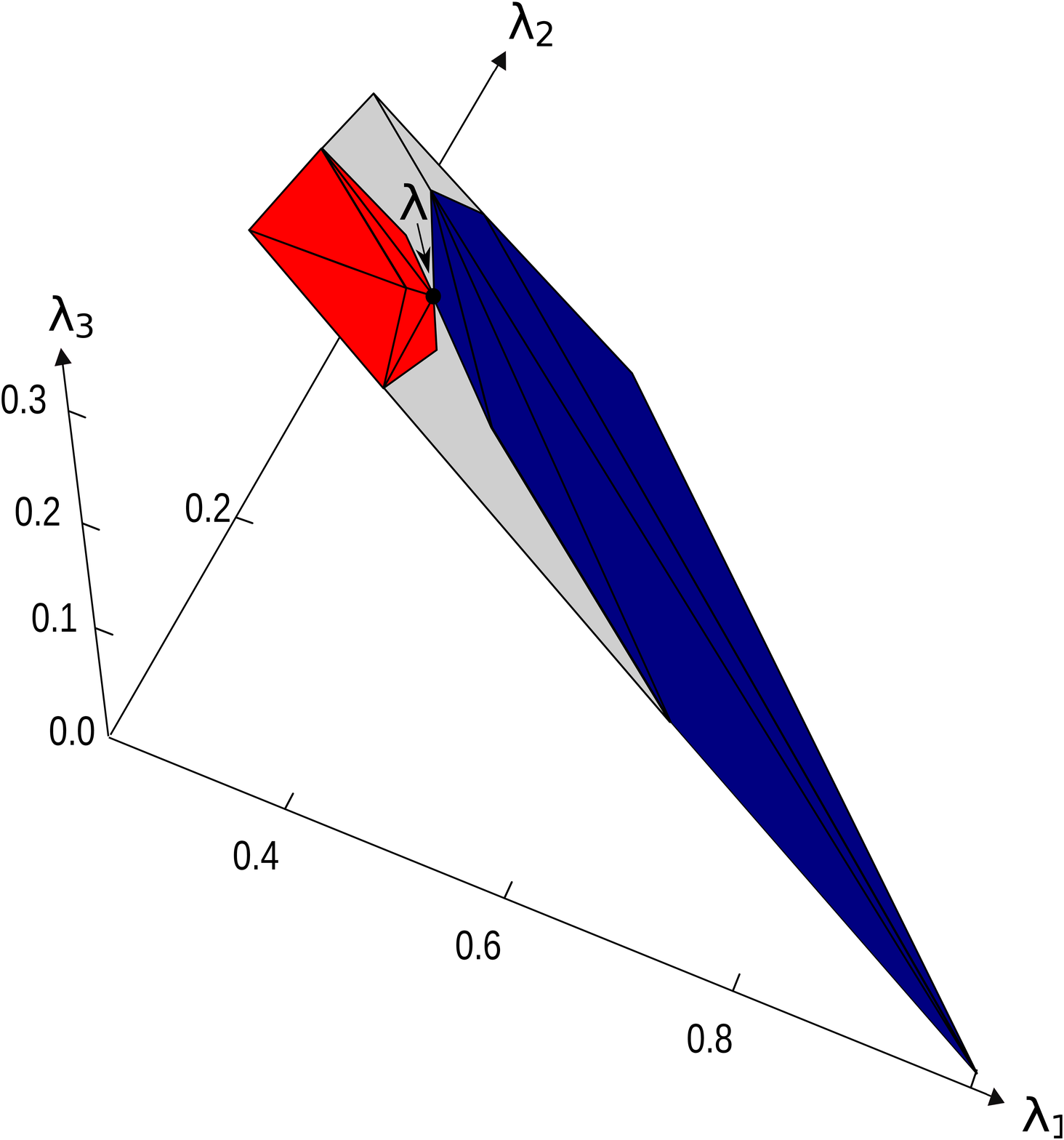}
 \caption{(colors online) The source and accessible set of a state of two four-level systems with sorted Schmidt vector $\lambda
 = (0.4,0.3,0.2,0.1)$. The valid, i.e. normalized, Schmidt vectors correspond to all states in the set
 $\{(\lambda_1,\lambda_2,\lambda_3) \ \mbox{s.t.} \ \lambda_i \geq 0, \sum_{i=1}^3 \lambda_i \leq 1\}$. The
 last coefficient is given by normalization. The convex set indicated by the shaded region (including the red and the blue region) highlights
 the set of sorted Schmidt vectors that is in one-to-one correspondence to the LU-equivalence classes. The red (blue) regions depict the
 source (accessible) set of the quantum state respectively. The vertices of the accessible set and its volume have been computed with the algorithm presented in \cite{avis2} 
 (see also Sec. \ref{sec:Accessible}). We obtain $E_s(\psi) = 0.904$ and $E_a(\psi) = 87/125 \approx 0.696$ for the depicted state.}
 \label{fig:4x4sets}
\end{figure}

Finally, we also consider quantum states of two four-level systems. Also for such a system we can depict the source and
accessible set of a given state (see Fig. \ref{fig:4x4sets}), as they are at most three-dimensional. We could not give an analytic formula for the 
accessible entanglement. However, the algorithm described in \cite{avis2} (see also Section \ref{sec:Accessible}) can be used 
to find the vertices and the volume of the accessible set of a given two-qutrit state. We used this algorithm to determine these 
properties for the state depicted in Fig. \ref{fig:4x4sets}. 
Analogously to before, the source entanglement can be computed using Lemma \ref{lem:SourceEnt}. It is given by
\begin{align*}
E_s(\psi) = & 4 \lambda_2^3+12 \lambda_2^2 \lambda_3-24 \lambda_2^2 \lambda_4-24 \lambda_2 \lambda_3^2+24\lambda_2 \lambda_3 \lambda_4 \nonumber\\
& +12 \lambda_2 \lambda_4^2-20 \lambda_3^3+12 \lambda_3^2 \lambda_4+18 \lambda_3^2 + 48 \lambda_3 \lambda_4^2 \nonumber \\
& -36 \lambda_3 \lambda_4+20 \lambda_4^3-30 \lambda_4^2+12 \lambda_4. \nonumber
\end{align*}

\section{Entanglement of multipartite systems}

In the previous section we showed that one can obtain explicit formulae for a whole class of new operational entanglement measures that 
characterize how easy it is to generate a single copy of a quantum state from a single copy of another quantum state via LOCC. 
Moreover, in \cite{us} we proved that these measures can also be used to completely characterize three-qubit entanglement.
Here, we will determine the source and accessible entanglement of generic four-qubit states.

Clearly the source and accessible volume can only be computed if one knows all the possible LOCC transformations of the states of interest. Note that in contrast to the bipartite case, the characterization of all possible LOCC transformations is generally very difficult, as also protocols involving infinitely many rounds of communication have to be considered \cite{Chitambar}. However, in this section we will derive the necessary and sufficient conditions for the existence of a LOCC transformation among pure states by completing the results on deterministic state transformations of four-qubit states presented in \cite{mes}. This criterion allows us then to compute the two entanglement measures.  Note that we want to consider deterministic transformations among fully entangled states, which are only possible among states within the same SLOCC class. In \cite{verstraete} it was shown that there exist infinitely many SLOCC classes for four-qubit states. Here, we only consider generic four-qubit states, which belong to the SLOCC classes denoted by $G_{abcd}$ with representatives of the form \cite{verstraete}
    \begin{align}
      \ket{\Psi}_{seed} \hspace*{-0.1cm}
      &= \hspace*{-0.1cm}\frac{a\hspace*{-0.1cm}+\hspace*{-0.1cm}d}{2} \left( \ket{0000}\hspace*{-0.1cm}+\hspace*{-0.1cm}\ket{1111} \right)
      + \frac{a\hspace*{-0.1cm}-\hspace*{-0.1cm}d}{2} \left( \ket{0011}\hspace*{-0.1cm}+\hspace*{-0.1cm}\ket{1100} \right) \nonumber \\
      &
      + \frac{b\hspace*{-0.1cm}+\hspace*{-0.1cm}c}{2} \left( \ket{0101}\hspace*{-0.1cm}+\hspace*{-0.1cm}\ket{1010} \right)
      + \frac{b\hspace*{-0.1cm}-\hspace*{-0.1cm}c}{2} \left( \ket{0110}\hspace*{-0.1cm}+\hspace*{-0.1cm}\ket{1001} \right)
      ,
      \label{eq:Gabcd}
    \end{align}
    with $b,c,d\in\C, \ a \in \R$ and $a^2 + \abs{b}^2+\abs{c}^2+\abs{d}^2=1$, $ a^2 \neq b^2,c^2,d^2,  \  b^2 \neq c^2 \neq d^2 \neq b^2$ 
    and the parameters fulfill the condition that there exists no $q \in \C\setminus\{1\}$ such that 
    $\lbrace a^2,b^2,c^2,d^2 \rbrace = \lbrace q a^2, q b^2, q c^2, q d^2 \rbrace$. These states are referred to as the \emph{seed states}, which are 
    parametrized by six real parameters, due to normalization and the global phase. Note that 
    $P_{i j} \ket{\Psi}_{seed} = \sigma_x^i \otimes \sigma_x^j \ket{\Psi}_{seed}$ for any particle permutation $P_{i j}$ with
    $i \neq j$, $i, j \in \lbrace 1,2,3,4 \rbrace$. That is, any seed state has the property that permuting the particles leads to a LU-equivalent state.
Any state in some SLOCC class $G_{abcd}$ can be written as $\ket{\Psi(g)}\equiv g \ket{\Psi}_{seed}$, with a local invertible operator (not necessarily 
determinant 1) $ g \in \mathcal{G}$, i.e. $g= g^1 \otimes g^2 \otimes g^3 \otimes g^4$, $g^i \in GL(2)$, and the corresponding seed state 
$\ket{\Psi}_{seed}$. Ignoring for the moment the normalization, the positive operators $G^i = (g^i)^{\dagger} g^i$ fulfill without loss of generality 
$\tr (G^i)=1$. Furthermore, they are positive full-rank operators written as 
$G^i = 1/2 \mathbb{1} + \sum_k \gamma_k^i \sigma_k$, with $\gamma_k^i \in \R$ and $0 \leq \abs{\boldsymbol{\gamma}^i} < 1/2$, 
where $\boldsymbol{\gamma}^i = (\gamma_1^i, \gamma_2^i, \gamma_3^i)$. Sometimes we will also use the notation 
$\gamma_w^i$ with $w \in \lbrace x,y,z \rbrace$ for the state parameters. Note that the symmetries of the seed states are given by 
$\lbrace \sigma_i^{\otimes 4}\rbrace_{i=0}^3$, i.e. $\sigma_i^{\otimes 4}\ket{\Psi}_{seed}=\ket{\Psi}_{seed}$, for all $i$. These symmetries can only simultaneously change the sign of two parameters
$\gamma_k^i$ and $\gamma_l^i$, with $k,l \in \lbrace 1,2,3 \rbrace, \ \ k \neq l$, for all $i$.  Thus, $G^i$ can be made unique \cite{mes}. 
By choosing $g^i = \sqrt{G^i}$ and sorting the parameters of the seed states the form
\begin{align}
\ket{\Psi (g^1,g^2,g^3,g^4)} = g \ket{\Psi}_{seed}
\label{standardForm}
\end{align}
 is a unique standard form of the state $\ket{\Psi(g^1,g^2,g^3,g^4)}$. Therefore, generic four-qubit states are in the same LU-equivalence class iff their 
 standard forms coincide. As we only consider representatives of LU-equivalence classes for the computation of the source and accessible volume we will 
 pick the state $\ket{\Psi(g^1,g^2,g^3,g^4)}$ as such. \\
 In \cite{mes} the \textit{maximally entangled set} (MES) for the generic four-
 qubit states has been derived. The MES is the minimal set of states from which any other genuinely multipartite entangled state can be obtained via LOCC.
 It has been shown that this set is of full measure for four-qubit states, in contrast to two- and three-qubit states. The reason for that is that most 
 four-qubit states are isolated, meaning that they can neither be reached nor converted into any other non LU-equivalent state. Hence, for isolated states 
 the source and accessible volume vanish. The non-isolated states have been shown to be (up to permutations of the local operators) \cite{mes}
 \begin{align}
\ket{\Psi (g^1,g_w^2,g_w^3,g_w^4)} = g^1 \otimes g_w^2 \otimes g_w^3 \otimes g_w^4 \ket{\Psi}_{seed},
\label{nonisolatedstates}
\end{align}
with $g_w^i \in \text{span}\lbrace \mathbb{1}, \sigma_w \rbrace$, such that $G_w^i = (g_w^i)^{\dagger} g_w^i = 1/2 \mathbb{1} + \gamma_{w}^i \sigma_{w}$, $w \in \lbrace x, y, z\rbrace$ and arbitrary $g^1$ as defined above. Note that we will in the following choose the two parameters $\gamma_1^1$ and $\gamma_2^1$ nonnegative and allow for negative values of the third parameter $\gamma_3^1$, i.e. $\gamma_1^1,\gamma_2^1 \in [0,1/2)$, $\gamma_3^1 \in (-1/2,1/2)$. We can define the standard form of $\ket{\Psi (g^1,g_w^2,g_w^3,g_w^4)}$ like that, as the symmetries of the seed state can always change the sign of two parameters of $\boldsymbol{\gamma}^1$ simultaneously as mentioned above. Hence, the non-isolated states form a 12-parameter family. As only this zero measure set of states allows for non-trivial LOCC transformations these are the only states we need to consider here.
Note that we consider here states up to permutations of the parties as they do not alter the necessary and sufficient conditions for LOCC convertibility and hence, we can easily compute the source and accessible volume of the permuted states (see below). The non-isolated states in the MES, that can not be reached by any other state but can access other states via LOCC, form a 10-parameter family with $g = g_w^1 \otimes g_w^2 \otimes g_w^3 \otimes g_w^4$, $w \in \lbrace x, y, z\rbrace$ (excluding $g_w^i \not\propto \mathbb{1}$ for exactly one $i$) \cite{mes}.

In the following we will derive the necessary and sufficient conditions for LOCC transformations among generic four-qubit states. Subsequently, we will always denote the state of interest by $\ket{\Psi( g^1,g^2,g^3,g^4)}$ as in Eq.\ \eqref{standardForm} and the states in the source or accessible set of the state of interest by $\ket{\Phi(h^1,h^2,h^3,h^4)} = h \ket{\Psi}_{seed}$. Note that one obviously only has to consider non-isolated states. First, one can easily show that LOCC transformations exist only among very particular pairs of states, as stated in the following observation which is proven in Appendix C.
\begin{observation}
A state  $\ket{\Psi(g)}$ can be transformed to a reachable state $\ket{\Phi(h^1,h_w^2,h_w^3,h_w^4)}$ by LOCC only if $g = g^1 \otimes g_w^2 \otimes g_w^3 \otimes g_w^4$, with the same $w \in \lbrace x,y,z \rbrace$.
\label{Observation}
\end{observation}
Note that not any state of the form $\ket{\Phi(h^1,h_w^2,h_w^3,h_w^4)}$ is reachable, as stated in the next lemma, which has been proven in \cite{mes}.
\begin{lemma}
A generic non-isolated four-qubit state $\ket{\Phi(h^1,h_w^2,h_w^3,h_w^4)} = h \ket{\Psi}_{seed}$ can be reached via LOCC by some other state $\ket{\Psi(g^1,g_w^2,g_w^3,g_w^4}$, with $w \in \lbrace x,y,z \rbrace$ iff (up to permutations) either
\begin{enumerate}[(i)]
\item $h = h^1 \otimes h_w^2 \otimes h_w^3 \otimes h_w^4$, with $h^1 \neq h^1_w$, and at least one $i\in\{2,3,4\}$ such that $h_w^i\not\propto \mathbb{1}$ and $h \neq h_v^1 \otimes h_w^2 \otimes \mathbb{1} \otimes \mathbb{1}$ for $v \neq w$, $v  \in \lbrace x,y,z \rbrace$ or
\item $h = h_v^1\otimes h_w^2 \otimes \mathbb{1} \otimes \mathbb{1}$, with $h_w^2 \not\propto \mathbb{1}$, $h_v^1 \not\propto \mathbb{1}$ and $v \neq w$, $v  \in \lbrace x,y,z \rbrace$, or
\item $h = h^1 \otimes \mathbb{1} \otimes \mathbb{1} \otimes \mathbb{1}$, with $h^1 \not\propto \mathbb{1}$.
\end{enumerate}
\label{Lemma1}
\end{lemma}
As before we call $\ket{\Phi(h^1,h_w^2,h_w^3,h_w^4)}$ with $h$ defined as in one of the three above cases a \textit{reachable} state.
Note that the reachable states are of course the non--isolated states excluding the non--isolated states in the MES. 

Let us now show from which states these states are reachable. In order to do so, let us denote by $\gamma_k^i$ the real state parameters of $\ket{\Psi(g^1,g_w^2,g_w^3,g_w^4)}$, e.g. $G^1 = (g^1)^{\dagger} g^1 = 1/2 \mathbb{1} + \sum_k \gamma_k^1 \sigma_k$ and by $\zeta_k^i$ the state parameters of $\ket{\Phi(h^1,h_w^2,h_w^3,h_w^4)}$ (with corresponding parameter vectors $\boldsymbol{\gamma}^i$ and $\boldsymbol{\zeta}^i$ respectively). Furthermore, we will use the notation $\gamma_{u(v)}^1$ (and the same for $\zeta_{u(v)}^1$) whenever we consider both the $u$ and $v$ component of $\boldsymbol{\gamma}^1$ for $u \neq v$, $u,v, \in \lbrace 1,2,3 \rbrace$. 
\begin{lemma} \label{lemma5}
A state $\ket{\Psi(g^1,g_w^2,g_w^3,g_w^4)}$ can non-trivially access a reachable state $\ket{\Phi(h^1,h_w^2,h_w^3,h_w^4)}$ within case $(I)$ of Lemma \ref{Lemma1} iff the following $I^{th}$ condition (for reachable states) holds for $I\in \{i,ii,iii\}$. 
\begin{enumerate}[(i)]
\item $\zeta_w^i\neq 0$ for \begin{enumerate}[(a)]
\item at least two $i\in \{1,2,3,4\}$, $\gamma_w^j = \zeta_w^j$ for $j = 1, 2,3,4$, and $\gamma_{v(u)}^1 = (2p-1) \zeta_{v(u)}^1$ with $p \in [1/2,1)$ and $\lbrace v,w,u \rbrace = \lbrace 1, 2, 3\rbrace$.
\item exactly one $i \in \lbrace 2,3,4 \rbrace$ and $\zeta_w^1 =0$, $\zeta_{v(u)}^1 \neq 0$ for $\lbrace v,w,u \rbrace = \lbrace 1, 2, 3\rbrace$ and either the conditions from above are fulfilled or $\gamma_w^k=0$ for all $k \neq i$, $k \in \lbrace 1,2,3,4 \rbrace$, $\gamma_{v(u)}^1 = 0$ and $\gamma_w^i \leq \zeta_w^i$.
\end{enumerate}
\item $\gamma_w^3 =\gamma_w^4 =0$,  $\gamma_{w(u)}^1 = 0$,  $\gamma_w^2  \leq \zeta_w^2$ and $\gamma_v^1 \leq \zeta_v^1$, with $\lbrace v,w,u \rbrace = \lbrace 1, 2, 3\rbrace$.
\item $\gamma_w^i =0$ for $i = 2,3,4$ and either \begin{enumerate}[(a)]
\item $\zeta_k^1 \neq 0$ $ \forall k \in \lbrace 1,2,3\rbrace$. Then with $r_k = \gamma_k^1 / \zeta_k^1$ the state parameters have to fulfill the three inequalities  $1 + r_1 - r_2 - r_3 \geq 0$, $1 -r_1 + r_2 - r_3 \geq 0$ and $1 - r_1 - r_2 + r_3 \geq 0$, with $0 \leq r_{1(2)} \leq 1$ and $-1 \leq r_3 \leq 1$.
\item Or $\zeta_k^1 =\gamma_k^1 = 0 $ for exactly one $k$. Then $\zeta_i ^1\geq \gamma_i^1 $ and $\zeta_j^1 \geq \gamma_j^1$ with $\lbrace i,j,k\rbrace = \lbrace 1,2,3 \rbrace$.
\item Or $\zeta_k^1 \neq 0$ and thus, $\gamma_k^1 \neq 0$ for exactly one $k$. Then $\zeta_k^1 \geq \gamma_k^1$ with $k \in  \lbrace 1,2,3 \rbrace$.
\end{enumerate}
\end{enumerate}
\label{Lemma2}
\end{lemma}

The proof of Lemma \ref{lemma5} can be found in Appendix C.
Note that Observation \ref{Observation}  together with Lemma \ref{Lemma1} and Lemma \ref{Lemma2} constitute the necessary and sufficient conditions for 
LOCC convertibility among generic four-qubit states. We will use them in the subsequent subsections to measure the entanglement contained in these states 
and summarize them now in the following table. We choose from now on without loss of generality $w=x$. \begin{table}[H]\small{
\begin{tabular}{| @{}m{2.45cm} | @{}m{2.5cm}| @{}m{3cm} | @{}m{0.7cm} |} 
\hline 
\textbf{Initial state} & \textbf{Final state} & \textbf{Necessary and sufficient conditions} & \textbf{Case}\\ 
\hline 
$\ket{\Psi(g^1,g_x^2,g_x^3,g_x^4)}$ & $\ket{\Phi(h^1,g_x^2,g_x^3,g_x^4)}$ & $\gamma_1^1 = \zeta_1^1$ and $\exists p \in [1/2,1)$  s.t. $(2p-1) \zeta_{2(3)}^1 = \gamma_{2(3)}^1$ & (ia)\\ 
\hline 
$\ket{\Psi(g_{y(z)}^1,g_x^2,\mathbb{1},\mathbb{1})}$ & $\ket{\Phi(h_{y(z)}^1,h_x^2,\mathbb{1},\mathbb{1})}$ & $\gamma_1^2 \leq \zeta_1^2$ and $\gamma_{2(3)}^1 \leq \zeta_{2(3)}^1$  & (ii)\\ 
\hline 
$\ket{\Psi(g^1,\mathbb{1},\mathbb{1},\mathbb{1})}$ & $\ket{\Phi(h^1,\mathbb{1},\mathbb{1},\mathbb{1})}$ & $g^1 \not\propto \mathbb{1}, g_x^1$, $\lbrace i,j,k \rbrace = \lbrace 1,2,3 \rbrace$ and either  \hspace*{-0.5cm}\begin{minipage}{3.2cm} \begin{itemize}
\item $\zeta_i^1 \neq 0$ and $1 - r_i + r_j -r_k \geq 0$ with $r_i = \gamma_i^1 / \zeta_i^1$, 
\item or $\zeta_k^1 = 0$ for exactly one $k$, $\gamma_k^1 = 0$, $\zeta_i^1 \geq \gamma_i^1$ and $\zeta_j^1 \geq \gamma_j^1$, 
\item or $\zeta_k^1 \neq 0$ and $\gamma_k^1 \neq 0 $ for exactly one $k$ and $\zeta_k^1 \geq \gamma_k^1$. 
\end{itemize} \end{minipage} & (iiia)  \normalcr[1.1cm] (iiib) \normalcr[1.2cm] (iiic)\\ 
\hline 
\multirow{ 2}{*}{$\ket{\Psi(g_x^1,\mathbb{1},\mathbb{1},\mathbb{1})}$ } & $\ket{\Phi(h^1,\mathbb{1},\mathbb{1},\mathbb{1})}$ & $\zeta_1^1 \geq \gamma_1^1$ & (iiia) \\ 
  & $\ket{\Phi(h_x^1,h_{y,z}^2,\mathbb{1},\mathbb{1})}$ & $\zeta_1^1 \geq \gamma_1^1$ & (ib) \\ 
\hline 
\multirow{ 2}{*}{$\ket{\Psi}_{seed}$ } & $\ket{\Phi(h^1,\mathbb{1},\mathbb{1},\mathbb{1})}$ &  & (iiia) \\ 
  & $\ket{\Phi(h_w^1,h_{u,v}^2,\mathbb{1},\mathbb{1})}$ &  & (ib) \\ 
\hline 
\end{tabular}} \caption{Summary of all possible nontrivial LOCC transformations of generic four-qubit states (up to permutations) with appropriate necessary and sufficient conditions (see Lemma \ref{Lemma2}). Note that also the normalization condition on the parameter vector, i.e. $0 \leq \abs{\boldsymbol{\gamma}^1}, \abs{\boldsymbol{\zeta}^1} < 1/2$ has to be fulfilled together with the necessary and sufficient conditions in the third column of the table. In the fourth column of the table the corresponding cases of Lemma \ref{Lemma2} are given for each of the LOCC conversions.} 
 \label{table1}
 \end{table}

Due to the fact that the criterion for the existence of a LOCC protocol does not depend on the seed parameters, it is evident that the measures will only depend on the state parameters and not on the seed parameters. To be more precise, we denote by $\ket{\Psi(a,b,c,d)}_{seed}$ the seed state given in Eq. (\ref{eq:Gabcd}). Then, the source and accessible entanglement of a state $g^1 \otimes g^2 \otimes g^3 \otimes g^4 \ket{\Psi(a,b,c,d)}_{seed}$ coincides with the corresponding entanglement of the state $g^1 \otimes g^2 \otimes g^3 \otimes g^4 \ket{\Psi(\tilde{a},\tilde{b},\tilde{c},\tilde{d})}_{seed}$ for any choice of the seed parameters $a,b,c,d$ and $\tilde{a},\tilde{b},\tilde{c},\tilde{d}$. 

\subsection{The source and accessible volume of generic four-qubit states}
\begin{figure}[H]
\centering
\includegraphics[width=0.35\textwidth]{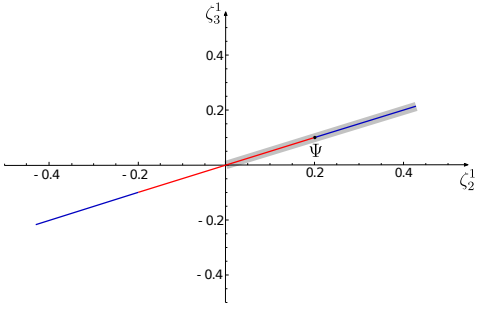}
\caption{(color online) Source (red line) and accessible (blue line) sets,
$M_s$ and $M_a$ (see Eqs.\ \eqref{Vagen} and \eqref{sourceVg} for the volumes), of a generic 4-qubit accessible state $\ket{\Psi(g^1,g_x^2,g_x^3,g_x^4)}$ with parameters $\gamma_1^1=0.15$, $\gamma_2^1=0.2$ and $\gamma_3^1=0.1$. The shaded region corresponds to valid states in the standard form which are in one-to-one correspondence with LU-equivalence classes.}
\label{figVaVs1}
\end{figure}

Given the necessary and sufficient conditions for LOCC convertibility of the previous section,  we will now compute the source and accessible volume of a 
state $\ket{\Psi(g^1,g_w^2,g_w^3,g_w^4)} = g \ket{\Psi}_{seed}$. Recall that we consider only fully entangled four-qubit states in the source and 
accessible set of $\ket{\Psi(g^1,g_w^2,g_w^3,g_w^4)}$. Note furthermore that we will use the freedom of choosing different measures $\mu$ for computing 
the source and accessible volume, as explained in Sec. \ref{sec:AccandSourc}. Whenever, the source or accessible states of certain states live in manifolds of different dimensionality, we choose different measures to compute their volumes (see also \cite{us}). Otherwise, we would assign zero values to the volumes of some states, even though they can be reached or accessed by other states. Hence, by using different measures for the computation of the volumes we can compare the relative strength of states whose volumes have the same dimension and moreover, regard e.g. states with a two-dimensional accessible volume obviously as infinitely more powerful than states with a one-dimensional accessible volume.
 We choose, as stated before, without loss of generality $w = x$, i.e. $\ket{\Psi(g^1,g_x^2,g_x^3,g_x^4)}= g^1 \otimes g_x^2 \otimes g_x^3 \otimes g_x^4 \ket{\Psi}_{seed}$, as all the other cases can be treated analogously. We will in the following consider the computation of the volumes of states, for which different necessary and sufficient conditions have to hold, in the same order as in Table \ref{table1}. Hence, we start by computing the source and accessible volume of a generic state $\ket{\Psi(g^1,g_x^2,g_x^3,g_x^4)}$ with $\gamma_x^i\neq 0$ for at least two $i\in \{1,2,3,4\}$ (see Lemma \ref{Lemma2} [case (ia)]), which is not in the MES, i.e. $g^1 \neq g_x^1$.  As before we denote by $\gamma_k^1$ the parameters of $g^1$ and by $\zeta_k^1$ the parameters of $h^1$.
Due to Lemma \ref{Lemma2} [case (ia)] the accessible set of parameters $\boldsymbol{\zeta}^1=(\zeta_1^1,\zeta_2^1,\zeta_3^1)$ is given by
\begin{align}
  \{ \boldsymbol{\zeta}^1 \ \text{s.t.} \ & \abs{\boldsymbol{\zeta}^1}<1/2
  \ , \ \zeta_1^1 = \gamma_1^1, \label{Vacond} \\ & \text{and} \ \exists  p, \ 1/2 \leq p <1 \ \text{s.t.} \ (2p-1) \zeta_{2(3)}^1 = \gamma_{2(3)}^1
  \}.\nonumber
  \end{align}
  Note that the first condition on the norm of the vector $\boldsymbol{\zeta}^1$ is due to the fact that $H^1$ has to be positive semidefinite.
Therefore, the accessible volume of $\ket{\Psi(g^1,g_x^2,g_x^3,g_x^4)}$ with $\gamma_x^i\neq 0$ for at least two $i\in \{1,2,3,4\}$ is a one-dimensional
line as the two parameters $\zeta_2^1$ and $\zeta_3^1$ can only be increased by the same proportion (see Fig. \ref{figVaVs1}). 
Then from the conditions in Eq.\ \eqref{Vacond} we get for any reachable state that $\zeta_3^1 = \zeta_2^1 \gamma_3^1 / \gamma_2^1$ and 
$\gamma_2^1 < \zeta_2^1< \sqrt{(1/4 - (\gamma_1^1)^2)/(1+(\gamma_3^1/   \gamma_2^1)^2)}$ ($\gamma_2^1$ can be chosen nonnegative in the standard form).  
Hence, the one-dimensional volume is determined by a line integral which results in
\begin{align}
   V_a (\ket{\Psi(g^1,g_x^2,g_x^3,g_x^4)}) \hspace*{-0.1cm}= \hspace*{-0.1cm} \sqrt{1/4\hspace*{-0.1cm}-\hspace*{-0.1cm}(\gamma_1^1)^2}\hspace*{-0.1cm}-\hspace*{-0.1cm}\sqrt{(\gamma_2^1)^2\hspace*{-0.1cm}+\hspace*{-0.1cm}(\gamma_3^1)^2}.
   \label{Vagen}
\end{align}
Thus, with $V_a^{sup} = 1/2$ we get for the accessible entanglement $E_a (\ket{\Psi(g^1,g_x^2,g_x^3,g_x^4)}) = 2  V_a (\ket{\Psi(g^1,g_x^2,g_x^3,g_x^4)})$. Note that we obtain here and in the following the supremum of the accessible (source) volume by simply optimizing the corresponding volume, which is in this case given by Eq.\ \eqref{Vagen}, over the valid parameter space. 
Obviously, we get similar expressions for the accessible volume of the states given by permutations of the parties of $\ket{\Psi(g^1,g_x^2,g_x^3,g_x^4)}$, e.g. for the state $\ket{\Psi(g_x^1,g^2,g_x^3,g_x^4)}$ the accessible volume is the same as in Eq.\ \eqref{Vagen} with $\gamma_k^1$ replaced by $\gamma_k^2$ for $k = 1,2,3$.\\
Let us now compute the source volume of the state $\ket{\Psi(g^1,g_x^2,g_x^3,g_x^4)}$. Due to Lemma \ref{Lemma2} [case (ia)] the parameters of all states in the source set of $\ket{\Psi(g^1,g_x^2,g_x^3,g_x^4)}$ with $\gamma_x^i\neq 0$ for at least two $i\in \{1,2,3,4\}$ (the other cases are treated below) are elements of the set 
\begin{align}
 \{ \boldsymbol{\zeta}^1 \ \text{s.t.} \ & \gamma_1^1 = \zeta_1^1 \ \text{and} \ \exists p, \ 1/2 \leq p <1 \ \text{s.t.} \nonumber \\ &(2p-1) \gamma_{2(3)}^1 = \zeta_{2(3)}^1
  \}.
\end{align}
It is easy to see that $\abs{\boldsymbol{\zeta}^1} \leq \abs{\boldsymbol{\gamma}^1}$ and thus, $\abs{\boldsymbol{\zeta}^1}  < 1/2$ is fulfilled for all vectors $\boldsymbol{\zeta}^1$.
Hence, the source volume of $\ket{\Psi(g^1,g_x^2,g_x^3,g_x^4)}$ is again a one-dimensional line defined by $\zeta_2^1<\gamma_2^1$ and the other parameter is fixed by $\zeta_3^1 = \zeta_2^1 \gamma_3^1/ \gamma_2^1$. Therefore, we obtain the following source volume
\begin{equation}
  V_s(\ket{\Psi(g^1,g_x^2,g_x^3,g_x^4)})=  \sqrt{(\gamma_2^1)^2+(\gamma_3^1)^2}.
  \label{sourceVg}
\end{equation}
The corresponding source entanglement measure is with $V_s^{sup} = 1/ 2$ given by $E_s(\ket{\Psi(g^1,g_x^2,g_x^3,g_x^4)}) = 1 - 2 V_s(\ket{\Psi(g^1,g_x^2,g_x^3,g_x^4)})$.
Note again that we get similar expressions for the source volume of the states that are given by permutations of the parties of $\ket{\Psi(g^1,g_x^2,g_x^3,g_x^4)}$ by simply replacing $\gamma_k^1$ with $\gamma_k^i$ for $i \in \lbrace 2,3,4\rbrace$.
Let us now consider states of the form $\ket{\Psi(g_x^1,g_x^2,g_x^3,g_x^4)}$, where at least two state parameters are non--vanishing, i.e. non-isolated states in the MES that are not the seed state. For these states the last conditions of Lemma \ref{Lemma2} [case (ia)] are fulfilled by choosing $p=1/2$, as $\gamma_2^1 = \gamma_3^1=0$. Hence, a state in the MES can reach any state of the form $\ket{\Phi(h^1,g_x^2,g_x^3,g_x^4)}$, as long as
$\gamma_1^1 = \zeta_1^1 $ and $\abs{\boldsymbol{\zeta}^1}<1/2 $. Moreover, the local operator of each of the parties can be changed, that is also for instance the state $\ket{\Phi(g_x^1,g_x^2,h^3,g_x^4)}$ can be reached as long as the corresponding conditions mentioned above are satisfied. Hence, the accessible set of parameters is given by
\begin{align}
\left\{ \boldsymbol{\zeta}^i \ \text{s.t.} \  \abs{\boldsymbol{\zeta}^i}<1/2 \ \text{and} \  \gamma_1^i = \zeta_1^i  \forall \  i   \in  \lbrace 1,2,3,4 \rbrace \right\}.
\end{align}
Therefore, the accessible volume of  $\ket{\Psi(g_x^1,g_x^2,g_x^3,g_x^4)}$ is the sum of four discs of radius $R_i=\sqrt{1/4-(\gamma_1^i)^2}$, for $i   \in  \lbrace 1,2,3,4 \rbrace$ i.e.
\begin{align}
  V_a(\ket{\Psi(g_x^1,g_x^2,g_x^3,g_x^4)})= \pi( &( 1/4-(\gamma_1^1)^2 )+( 1/4-(\gamma_1^2)^2 )+\nonumber \\ &( 1/4-(\gamma_1^3)^2 )+( 1/4-(\gamma_1^4)^2 )).
  \end{align}
 It can be easily seen that the supremum of the accessible volume is given by $V_a^{sup} = \pi$. Thus, we obtain for the accessible entanglement of convertible states in the MES $E_a(\ket{\Psi(g_x^1,g_x^2,g_x^3,g_x^4)}) = ( 1/4-(\gamma_1^1)^2 )+( 1/4-(\gamma_1^2)^2 )+ ( 1/4-(\gamma_1^3)^2 )+( 1/4-(\gamma_1^4)^2 )$.
The source volume of states in the MES is zero, as no state can reach a state in the MES via LOCC, i.e. $V_s(\ket{\Psi(g_x^1,g_x^2,g_x^3,g_x^4)}) = 0$ and hence, the source entanglement is equal to one, i.e. $E_s(\ket{\Psi(g_x^1,g_x^2,g_x^3,g_x^4)}) = 1$.

\begin{figure}[H]
\centering
\includegraphics[width=0.28\textwidth]{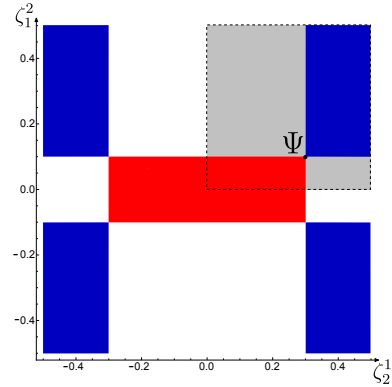}
\caption{(color online) Source (red rectangle) and accessible (blue rectangle) sets,
$M_s$ and $M_a$ (see Eqs.\ \eqref{Vagygx} and \eqref{Vsgygx} for the volumes), of a 4-qubit accessible state of the form $\ket{\Psi(g_y^1,g_x^2,\mathbb{1},\mathbb{1})}$ with parameters $\gamma_1^2=0.1$, $\gamma_2^1=0.3$. The dashed line encloses the grey area of valid states in the standard form corresponding to LU-equivalence classes. Hence, only states inside this grey area, where each of them is a representative of one LU-equivalence class, are in the source/accessible set.}
\label{VsVa2}
\end{figure}

Let us now consider the source and accessible volume of states as in Lemma \ref{Lemma1} [case (ii)]. Here, we choose without loss of generality states of 
the form $\ket{\Psi(g_y^1,g_x^2,\mathbb{1},\mathbb{1})} = g_y^1 \otimes g_x^2 \otimes \mathbb{1} \otimes \mathbb{1} \ket{\Psi}_{seed}$.
Due to Lemma \ref{Lemma2} these states can only be transformed into states of the form $\ket{\Phi(h_y^1,h_x^2,\mathbb{1},\mathbb{1})}$ 
(see also Table \ref{table1}, second row). The necessary and sufficient conditions for deterministic transformations into such a state are given in Lemma \ref{Lemma2} [case (ii)], which imply that the accessible set of parameters is 
\begin{equation}
\left\{(\zeta_2^1, \zeta_1^2) \ \text{s.t.} \ \gamma_2^1 < \zeta_2^1 \ \text{and} \ \gamma_1^2 < \zeta_1^2 \right\}.
\end{equation}
 Hence, the accessible volume is equal to the area of a rectangle (see Figure \ref{VsVa2}), i.e.
\begin{equation}
V_a(\ket{\Psi(g_y^1,g_x^2,\mathbb{1},\mathbb{1})}) =  (1/2- \gamma_2^1) (1/2- \gamma_1^2).
\label{Vagygx}
\end{equation}
Then, with $V_a^{sup} = 1/4$ the corresponding entanglement measure is equal to $E_a(\ket{\Psi(g_y^1,g_x^2,\mathbb{1},\mathbb{1})}) = 4 V_a(\ket{\Psi(g_y^1,g_x^2,\mathbb{1},\mathbb{1})})$.
For the source volume of $\ket{\Psi(g_y^1,g_x^2,\mathbb{1},\mathbb{1})}$ the conditions on the parameters of states of the form $\ket{\Phi(h_y^1,h_x^2,\mathbb{1},\mathbb{1})}$ in the source set are given by
\begin{equation}
\left\{(\zeta_2^1, \zeta_1^2) \ \text{s.t.} \ \gamma_2^1 > \zeta_2^1 \ \text{and} \ \gamma_1^2 > \zeta_1^2 \right\},
\end{equation}
and thus, the volume is again equal to the area of a rectangle (Figure \ref{VsVa2}), i.e.
\begin{equation}
V_s (\ket{\Psi(g_y^1,g_x^2,\mathbb{1},\mathbb{1})})= 4 \gamma_2^1 \gamma_1^2.
\label{Vsgygx}
\end{equation}
The source entanglement is given by $E_s (\ket{\Psi(g_y^1,g_x^2,\mathbb{1},\mathbb{1})}) = 1- V_s (\ket{\Psi(g_y^1,g_x^2,\mathbb{1},\mathbb{1})})$, as $V_s^{sup} = 1$.

Next we examine the source and accessible volume of states of the form $\ket{\Psi(g^1,\mathbb{1},\mathbb{1},\mathbb{1})}$ as in Lemma \ref{Lemma1} [case (iii)] (see also Table \ref{table1} third row). Let us assume for now that none of the state parameters is vanishing, i.e. $\gamma_i^1 \neq 0$ and $\zeta_i^1 \neq 0$, $i \in \lbrace 1,2,3 \rbrace$. The necessary and sufficient conditions for possible LOCC transformations of these states  already show, that the corresponding volumes are always three-dimensional, as one can vary all three $\gamma_i^1$ parameters in accordance with the three inequalities given in Lemma \ref{Lemma2} [case (iiia)]. Hence, all states $\ket{\Phi(h^1,\mathbb{1},\mathbb{1},\mathbb{1})}$ are in the accessible set of  $\ket{\Psi(g^1,\mathbb{1},\mathbb{1},\mathbb{1})}$ if the parameters fulfill \begin{align}
\lbrace \boldsymbol{\zeta}^1  \ \text{s.t.} \ & \abs{\boldsymbol{\zeta}^1}<1/2 \ \text{and} \  1\hspace*{-0.1cm} +\hspace*{-0.1cm} r_i \hspace*{-0.1cm}-\hspace*{-0.1cm} r_j \hspace*{-0.1cm}-\hspace*{-0.1cm} r_k \hspace*{-0.1cm} \geq \hspace*{-0.1cm} 0 \nonumber \\ & \text{for} \ \lbrace i,j,k \rbrace = \lbrace 1,2,3 \rbrace  \rbrace,
\label{condVa}
\end{align}
where $ r_l = \gamma_l^1 / \zeta_l^1$. Note that in case $g^1 = g_w^1$ with $w \in \lbrace x,y,z \rbrace$ additional LOCC transformations are possible, implying that more states are accessible (see below). For that reason we consider first the case $g^1 \neq g_w^1$ for any $w \in \lbrace x,y,z \rbrace$. With the help of cond.\ \eqref{condVa} one can compute the accessible volume of any given state of the form $\ket{\Psi(g^1,\mathbb{1},\mathbb{1},\mathbb{1})}$. It is, however, not easy to give a closed expression of the accessible volume for arbitrary parameters of $g^1$, as it is difficult to get the limits of the integral that gives the accessible volume from cond.\ \eqref{condVa}. However, given $g^1$ this volume can be easily computed. Note that for states with $\gamma_k^1=0$ for exactly one $k \in \lbrace 1,2,3 \rbrace$ also states with $\zeta_k^1=0$ can be reached, see Lemma \ref{Lemma2} [case (iiib)]. The part of the accessible volume containing only states with $\zeta_k^1=0$ is two dimensional (as $\zeta_k^1=0$) and hence, of measure zero, as long as the three-dimensional accessible volume exists. However, there are states $\ket{\Psi(g^1,\mathbb{1},\mathbb{1},\mathbb{1})}$ with $\gamma_k^1=0$ for which cond.\ \eqref{condVa} can not be fulfilled, meaning that they can only reach states with $\zeta_k^1=0$. Thus, for these states the accessible volume is only two-dimensional as explained before (see also below for the source volume of states with $\zeta_k^1 = 0$).

\begin{figure}[H]
\centering
\includegraphics[width=0.28\textwidth]{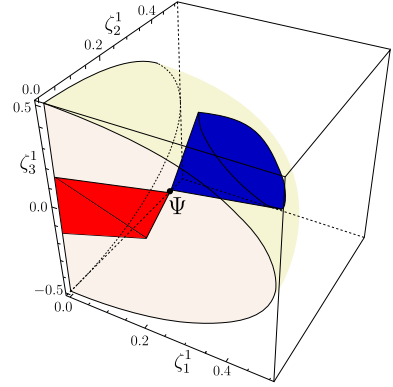}
\caption{(color online) Source (red polyhedron) and accessible (blue) sets,
$M_s$ and $M_a$ (see Eqs.\ \eqref{condVa} and \eqref{Vsg1} for the volumes), of a 4-qubit accessible state $\ket{\Psi(g^1,\mathbb{1},\mathbb{1},\mathbb{1})}$ 
with parameters $\gamma_1^1=0.23$, $\gamma_2^1=0.13$ and $\gamma_3^1= 0.15$. The volume of the quadrant of a sphere corresponds to a part of the 
accessible volume of the seed state $\ket{\Psi}_{seed}$.}
\label{figVsVa3}
\end{figure}

\begin{figure}[H]
\centering
\includegraphics[width=0.28\textwidth]{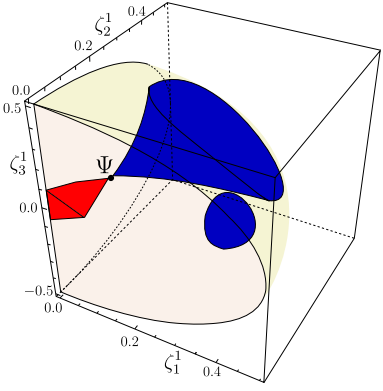}
\caption{(color online) Source (red polyhedron) and accessible (blue) sets,
$M_s$ and $M_a$, of a 4-qubit accessible state $\ket{\Psi(g^1,\mathbb{1},\mathbb{1},\mathbb{1})}$ with parameters $\gamma_1^1=0.09$, $\gamma_2^1=0.1$ 
and $\gamma_3^1= 0.08$. Interestingly, the accessible set contains two sets that are not connected.}
\label{figVsVa4}
\end{figure}

\begin{figure}[H]
\centering
\includegraphics[width=0.28\textwidth]{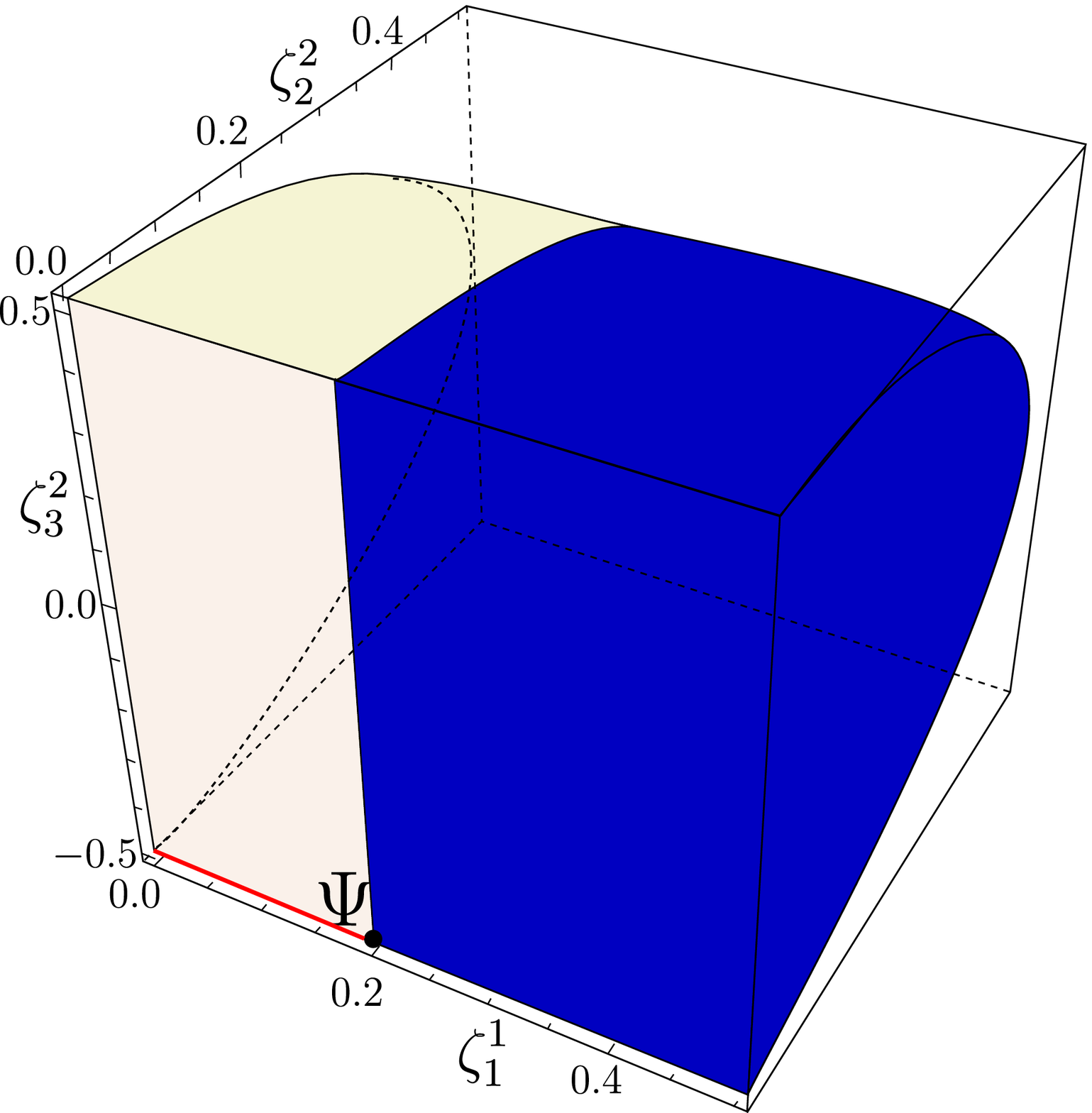}
\caption{(color online) A part of the accessible set (blue), $M_a^2$ (see Eq.\ \eqref{Va2} for the volume), of a four-qubit state $\ket{\Psi(g_x^1,\mathbb{1},\mathbb{1},\mathbb{1})}$ with $\gamma_1^1=0.2$. The volume is equal to $1/2$ the volume of a cylinder with height $(1/2-\gamma_1^1)$ and radius $1/2$. The red line is the source set of $\ket{\Psi(g_x^1,\mathbb{1},\mathbb{1},\mathbb{1})}$  (the source volume is one-dimensional). The volume of the larger set corresponds to a part of the cylinder belonging to the accessible volume of the seed state, for which $\gamma_1^1=0$.}
\label{figVsVa5}
\end{figure}

Let us now determine the source volume for states of the form $\ket{\Psi(g^1,\mathbb{1},\mathbb{1},\mathbb{1})}$ in terms of the state parameters.
We consider first the case where none of the coefficients of $g^1$ in the Pauli basis vanish. For all states $\ket{\Phi(h^1,\mathbb{1},\mathbb{1},\mathbb{1})}$ in the source set of $\ket{\Psi(g^1,\mathbb{1},\mathbb{1},\mathbb{1})}$ again the three inequalities $1 + r_i - r_j - r_k \geq 0$ for $\lbrace i,j,k \rbrace = \lbrace 1,2,3 \rbrace$ have to be fulfilled, where here $r_l = \zeta_l^1 / \gamma_l^1$ (Lemma\ \ref{Lemma2} [case (iiia)]). These inequalities also imply the necessary conditions $\abs{\gamma_l^1 } \geq \abs{\zeta_l^1}$ for $l \in \lbrace 1,2,3\rbrace$. 
Thus, it can be easily seen that the normalization condition on the parameter vector, i.e. $\abs{\boldsymbol{\zeta}^1} < 1/2$, is automatically fulfilled, 
as $\abs{\boldsymbol{\gamma}^1} < 1/2$ holds. We can compute the source volume of $\ket{\Psi(g^1,\mathbb{1},\mathbb{1},\mathbb{1})}$ by integrating over 
the valid parameter ranges for the $\zeta_l^1$ given by the three inequalities and obtain
\begin{align}
V_s (\ket{\Psi(g^1,\mathbb{1},\mathbb{1},\mathbb{1})}) &=  \int \hspace*{-0.2cm} \int \hspace*{-0.2cm}  \int_{R(\zeta_1^1,\zeta_2^1,\zeta_3^1)}\hspace*{-0.8cm}  d\boldsymbol{\zeta}^1 \nonumber \\ &= \int_0^1 \hspace*{-0.2cm} \int_0^{r_2}\hspace*{-0.2cm}  \int_{-1+r_1+r_2}^{1+r_1-r_2} \hspace*{-1.3cm}\gamma_1^1 \gamma_2^1 \abs{\gamma_3^1} dr_3 dr_1 dr_2  \nonumber \\ &+ \int_0^1 \hspace*{-0.2cm} \int_{r_2}^1 \hspace*{-0.1cm}  \int_{-1+r_1+r_2}^{1-r_1+r_2} \hspace*{-1.3cm} \gamma_1^1 \gamma_2^1 \abs{\gamma_3^1} dr_3 dr_1 dr_2 \nonumber \\  &=  2/3 \gamma_1^1 \gamma_2^1 \abs{\gamma_3^1}.
\label{Vsg1}
\end{align}
Note that out of mathematical convenience we made a change of variables in the integral above. 
Hence, the source entanglement is with $V_s^{sup} = 1/(36 \sqrt{3})$ given by $E_s(\ket{\Psi(g^1,\mathbb{1},\mathbb{1},\mathbb{1})})  = 1 - 24 \sqrt{3} \gamma_1^1 \gamma_2^1 \abs{\gamma_3^1}$.
In Fig. \ref{figVsVa3} we plot the source and accessible volume of a state $\ket{\Psi(g^1,\mathbb{1},\mathbb{1},\mathbb{1})}$ of the above form. Both volumes lie inside the quadrant of a sphere, which is a part of the accessible volume of the seed state, as explained below.
Furthermore, the source volume of states of the form $\ket{\Psi(g^1,\mathbb{1},\mathbb{1},\mathbb{1})}$ with vanishing parameters can be easily computed with the help of the conditions in Lemma \ref{Lemma2} [case (iiib) and case (iiic)]. If there is one parameter vanishing, i.e. without loss of generality $\gamma_1^1=0$, the two-dimensional source volume is given by $V_s(\ket{\Psi(g^1,\mathbb{1},\mathbb{1},\mathbb{1})}) = \gamma_2^1 \gamma_3^1$ with the source entanglement equal to $E_s(\ket{\Psi(g^1,\mathbb{1},\mathbb{1},\mathbb{1})}) = 1-4 \gamma_2^1 \gamma_3^1$. Moreover, the source volume of a state with two vanishing parameters, i.e. without loss of generality $\gamma_1^1 = \gamma_2^1 = 0$, is one-dimensional and reads $V_s(\ket{\Psi(g_z^1,\mathbb{1},\mathbb{1},\mathbb{1})}) = \gamma_3^1$ with $E_s(\ket{\Psi(g_z^1,\mathbb{1},\mathbb{1},\mathbb{1})}) = 1-2 \gamma_3^1$. \\
Note also that there exist states of the form $\ket{\Psi(g^1,\mathbb{1},\mathbb{1},\mathbb{1})}$ for which the accessible volume consists of two sets that are not connected (in the used parameterization). An example of such a state is given in Fig.\ \ref{figVsVa4}.
Now we consider the accessible volume of states with $g^1 = g_x^1\neq 0$ (we choose again $w=x$ without loss of generality) and the seed state (see Table\ \ref{table1} fourth and last row). As for the seed states additional LOCC transformations are possible, we have to treat them separately. Let us start with the accessible volume of states of the form $\ket{\Psi(g_x^1,\mathbb{1},\mathbb{1},\mathbb{1})}$, with $g_x^1 \not\propto \mathbb{1}$. We have to split the computation of the accessible volume of $\ket{\Psi(g_x^1,\mathbb{1},\mathbb{1},\mathbb{1})}$ into two parts, as these states can be transformed into states of different forms (Table\ \ref{table1} fourth row). First of all according to Lemma \ref{Lemma2} [case (iiia)] these states can be transformed via LOCC into all states of the form $\ket{\Phi(h^1,\mathbb{1},\mathbb{1},\mathbb{1})} $ fulfilling the two inequalities $1 \pm r_1 \geq 0$ \footnote{Note that in this case $r_2 = r_3 = 0 $.}. As we can choose $\gamma_1^1, \zeta_1^1 \geq 0$ the conditions above are equivalent to $\gamma_1^1 \leq \zeta_1^1$. Furthermore, the two parameters $\zeta_2^1, \zeta_3^1$ can be changed arbitrarily as long as they obey the condition on the norm, i.e. $\abs{\boldsymbol{\zeta}^1}<1/2$. Hence, the first part of the accessible volume of $\ket{\Psi(g_x^1,\mathbb{1},\mathbb{1},\mathbb{1})}$ is given by
\begin{align}
\label{eqVa1}
V_{a,1}(\ket{\Psi(g_x^1,\mathbb{1},\mathbb{1},\mathbb{1})})\hspace*{-0.1cm} &= \hspace*{-0.2cm}\int_{\hspace*{-0.1cm}\gamma_1^1}^{1/2}\hspace*{-0.3cm} \int_{-\sqrt{1/4-(\gamma_1^1)^2}}^{\sqrt{1/4-(\gamma_1^1)^2}}\hspace*{-0.1cm} \int_{0}^{\sqrt{1/4-(\gamma_1^1)^2-(\zeta_3^1)^2}}\hspace*{-1.7cm} d\zeta_2^1 d\zeta_3^1 d\zeta_1^1 \nonumber \\ & = 
\int_{\gamma_1^1}^{1/2} \int_0^{\pi} \int_0^{\sqrt{1/4-z^2}} r dr d\theta dz
\nonumber \\ & = 
\pi/2 \left(\frac{1}{12}-\frac{\gamma_1^1}{4}+\frac{(\gamma_1^1)^3}{3}\right),
\end{align}
where we converted the integral into cylindrical coordinates. 
Furthermore, the state $\ket{\Psi(g_x^1,\mathbb{1},\mathbb{1},\mathbb{1})}$ can be transformed via LOCC into all states of the form $\ket{\Phi(h_x^1, h_{y,z}^2,\mathbb{1},\mathbb{1})}$  (up to permutations), with $(h_{y,z}^2)^{\dagger} h_{y,z}^2 = 1/2 \mathbb{1} + \zeta_2^2 \sigma_2 + \zeta_3^2 \sigma_3$, see Lemma \ref{Lemma2} [case (ib)]. The transformation is done by first converting $\ket{\Psi(g_x^1,\mathbb{1},\mathbb{1},\mathbb{1})}$ into $h_x^1\otimes \mathbb{1}^{\otimes 3} \ket{\Psi}_{seed}$ and then converting this state into the final state $\ket{\Phi(h_x^1, h_{y,z}^2,\mathbb{1},\mathbb{1})}$. Thus, the necessary and sufficient conditions on the parameters are $\gamma_1^1 \leq \zeta_1^1$ (Lemma \ref{Lemma2} [case (ib)]) and $\zeta_2^2, \zeta_3^2$ fulfilling the normalization condition $\abs{\boldsymbol{\zeta}^2} < 1/2$ (with $\zeta_1^2 = 0$). This leads to the second part of the accessible volume given by
\begin{align}
V_{a,2}(\ket{\Psi(g_x^1,\mathbb{1},\mathbb{1},\mathbb{1})}) &= \int_{-1/2}^{1/2} \int_0^{\sqrt{1/4-(\zeta_3^2)^2}} \int_{\gamma_1^1}^{1/2} d\zeta_1^1 d\zeta_2^2 d\zeta_3^2‚ \nonumber\\ & = \frac{1}{16} \pi ( 1- 2 \gamma_1^1).
\label{Va2}
\end{align}
This part of the accessible volume of $\ket{\Psi(g_x^1,\mathbb{1},\mathbb{1},\mathbb{1})}$ is depicted in Fig.\ \ref{figVsVa5} for a random state.
Furthermore, all states given by permutations of the local operators of $\ket{\Phi(h_x^1, h_{y,z}^2,\mathbb{1},\mathbb{1})}$ (e.g. $\ket{\Phi(h_x^1,\mathbb{1}, h_{y,z}^2,\mathbb{1})}$) are also in the accessible set of $\ket{\Psi(g_x^1,\mathbb{1},\mathbb{1},\mathbb{1})}$ and thus, we have to count $V_{a,2}$ three times. Therefore, the total accessible volume is equal to
\begin{align}
V_a (\ket{\Psi(g_x^1,\mathbb{1},\mathbb{1},\mathbb{1})}) &= V_{a,1} + 3 V_{a,2} \nonumber \\ &= 1/48 \pi (11 + 8 \gamma_1^1 [ (\gamma_1^1)^2 -3] ).
\end{align}
 The accessible entanglement is then given by
$ E_a (\ket{\Psi(g_x^1,\mathbb{1},\mathbb{1},\mathbb{1})}) = 1 + 8/11 \gamma_1^1 [ (\gamma_1^1)^2 -3]$ as $V_a^{sup} = \frac{11}{48} \pi$.
Furthermore, we have to compute the source volume of a state of the form $\ket{\Psi(g_x^1, g_{y,z}^2,\mathbb{1},\mathbb{1})}$ (Lemma \ref{Lemma2} [case (ib)]), as this state can be reached on the one hand by states of the form $\ket{\Phi(h_x^1, \mathbb{1},\mathbb{1},\mathbb{1})}$ with $\zeta_1^1 \leq \gamma_1^1$ as we have just seen (Table\ \ref{table1} fourth row). Hence, the first part of the one-dimensional source volume of $\ket{\Psi(g_x^1, g_{y,z}^2,\mathbb{1},\mathbb{1})}$ is given by
\begin{equation}
V_{s,1}(\ket{\Psi(g_x^1, g_{y,z}^2,\mathbb{1},\mathbb{1})}) = \int_0^{\gamma_1^1} d\zeta_1^1 = \gamma_1^1.
\end{equation}
On the other hand according to Lemma \ref{Lemma2} [case (ib)] all states of the form $\ket{\Phi(g_x^1, h_{y,z}^2,\mathbb{1},\mathbb{1})}$ are also in the source set of $\ket{\Psi(g_x^1, g_{y,z}^2,\mathbb{1},\mathbb{1})}$ (see Table\ \ref{table1} first row). Thus, the second part of the source volume of $\ket{\Psi(g_x^1, g_{y,z}^2,\mathbb{1},\mathbb{1})}$ is given in Eq. \eqref{sourceVg}, where one simply has to replace $\gamma_{2(3)}^1$ with $\gamma_{2(3)}^2$.  The total source volume is obtained by taking the sum of both parts and reads
\begin{equation}
V_s (\ket{\Psi(g_x^1, g_{y,z}^2,\mathbb{1},\mathbb{1})}) = \gamma_1^1 + \sqrt{(\gamma_2^2)^2+(\gamma_3^2)^2}.
\end{equation}
As $V_s^{sup} = 1$ the source entanglement is given by $E_s(\ket{\Psi(g_x^1, g_{y,z}^2,\mathbb{1},\mathbb{1})}) = 1 - V_s(\ket{\Psi(g_x^1, g_{y,z}^2,\mathbb{1},\mathbb{1})})$. It is easy to see that the accessible volume of states of the form $\ket{\Psi(g_x^1, g_{y,z}^2,\mathbb{1},\mathbb{1})}$ is given by Eq. \eqref{Vagen} (see Lemma \ref{Lemma2} [case (ia)]), where one again has to replace $\gamma_i^1$ with $\gamma_i^2$ for $i \in \lbrace 1,2,3 \rbrace$ and set $\gamma_1^2 = 0$. 

Note that the seed state $\ket{\Psi}_{seed}$ has the maximum accessible volume, as we will show in the following. The seed state can be transformed up to permutations into either states of the form $\ket{\Phi(h^1,\mathbb{1},\mathbb{1},\mathbb{1})} $ with arbitrary $h^1$, see Lemma \ref{Lemma2} [case (iii)], or all states of the form $\ket{\Phi(h_w^1, h_{u,v}^2,\mathbb{1},\mathbb{1})}$ with $\lbrace u,v,w \rbrace = \lbrace x,y,z \rbrace$, see Lemma \ref{Lemma2} [case (ib)] (Table\ \ref{table1} last row). Hence, the three dimensional accessible volume of the seed state is equal to four times $1/4$ of the volume of a sphere with radius $1/2$ ($\abs{\boldsymbol{\zeta}^1}<1/2$ and $\zeta_1^1 >0, \ \zeta_1^2 >0$) plus 36 times ($\lbrace u,v,w \rbrace =  \lbrace x,y,z \rbrace$ and including all party permutations) 1/2 of the volume of a cylinder with radius $1/2$ and height $1/2$, see Fig.\ \ref{figVsVa5} ($\sqrt{(\zeta_u^2)^2+(\zeta_v^2)^2}<1/2$ and $0\leq \zeta_w^1 < 1/2$), i.e.
\begin{equation}
V_a ( \ket{\Psi}_{seed}) = \pi /6 + 9/4 \pi = 29/12 \pi.
\end{equation}
Thus, any seed state obtains the maximum value for the accessible entanglement, i.e. $E_a (\ket{\Psi}_{seed}) = 1$.
Summarizing the results on the accessible and source entanglement of generic four-qubit states we highlight again the three-dimensional accessible volume of the seed states, which is the maximum accessible volume of all generic states. This can already be seen in Table \ref{table1}, where in the last row for a conversion from the seed state to other states no additional necessary and sufficient conditions apart from the condition on the norm of the parameter vectors $\boldsymbol{\zeta}^1$ have to be fulfilled. Hence, the seed states can reach the most other states deterministically via LOCC. Furthermore, when considering transformations between states as in the first row of Table \ref{table1} the states in the MES, i.e. $\ket{\Psi(g_x^1,g_x^2,g_x^3,g_x^4)}$ have a two-dimensional accessible volume and they maximize obviously the source entanglement. Whereas, states that are not in the MES have in this case only a one-dimensional accessible volume and are therefore, infinitely less powerful than states in the MES.

\section{Conclusions}

In \cite{us} we have introduced two classes of entanglement measures for general multipartite quantum states: the accessible entanglement and the source 
entanglement. These measures have a clear operational meaning related to the relative usefulness of the state under single-copy deterministic LOCC 
manipulation. Moreover, whenever these transformations can be characterized, the measures can be computed. In \cite{us} we showed this in the case of 
three-qubit pure states and provided explicit formulae in these cases for both $E_a$ and $E_s$. In this article we have further demonstrated 
the applicability of our approach by analyzing the case of bipartite pure states of arbitrary dimension and of generic four-qubit pure states. 
In the first case, using tools of convex polytopes, we provided a closed expression for the source entanglement of an arbitrary pure bipartite state of 
Schmidt rank $d$ considering all possible dimensions on which $M_s$ can be supported, $\{E_s^k\}_{k\geq d}$.
The accessible volume turns out to be more complicated; however, there exist algorithms that 
allow to compute all measures $\{E_a^k\}_{k\leq d}$. Moreover, in the case of bipartite states of low dimension such as $d=2,3$ we have also obtained 
analytic expressions for the accessible entanglement. In the second case of four-qubit generic states, we have completed the analysis of \cite{mes} 
characterizing all possible LOCC conversions inside this class. Using this we have provided explicit formulae for $E_a$ and $E_s$ for all these states.

The results presented here show the versatility and applicability of the measures we introduced in \cite{us}, having covered the most relevant classes of 
pure few-body entangled states. We hope that having measures that are both computable and with a clear operational meaning will help to understand better
the properties and potential applications of multipartite quantum systems. For the future, our research opens several directions for further investigation.
First, it seems desirable to analyze in detail more many-body cases with the aim of obtaining closed expressions for our measures 
(or efficient algorithms for their computation). In this context it would also be interesting to exploit the results of \cite{mixed} to try to evaluate our
measures for mixed states. Moreover, we would like to connect our measures with applications. An interesting option is to study the role of our measures as 
figures of merit for known quantum information protocols. This could then lead to the identification of the most relevant multipartite states and maybe 
allow to devise new applications of multipartite entanglement. Also, we hope that our measures can bring new insights to understand the interplay between 
entanglement and many-body physics in, for example, phase transitions. Apart from that, the investigation of LOCC transformations among several copies of 
a given state and $\epsilon$-LOCC transformations \footnote{With $\epsilon$-LOCC transformations we refer to LOCC transformations in which a small 
deviation in the target state is allowed with respect to some suitably chosen distinguishability measure.} in this context will be relevant.\\

After completing this work we became aware of Ref. \cite{Postnikov}, where the volume of polytopes of the form of $\mathcal{M}^{LU}_s$ 
in Eq. (\ref{eq:Rado}) was calculated.
\begin{acknowledgments}
This research was funded by the Austrian Science Fund (FWF) Grant No. Y535-N16 and the Spanish MINECO
through grants MTM 2010-21186-C02-02 and MTM 2011-26912 and the CAM regional research consortium QUITEMAD+CM S2013/ICE-2801.
\end{acknowledgments}

\begin{appendix}

\section{The source volume of non-generic states}
\label{sec:AppendixA}

In this appendix we show that the formula for the source volume given in Eq. (\ref{eq:SourceVolume}), which was derived for states with
non-degenerate Schmidt coefficients, is valid in general. This derivation was based on the fact that $M_s^{LU}(\psi)$ is a simple polytope for a
generic state, such that
the formula in Eq. (\ref{Eq_Volume}) was applicable to compute its volume. However, $M_s^{LU}(\psi)$ fails to be simple in general for degenerate
states, i.e. states with at least two identical Schmidt coefficients. Consider,
e.g., the state with Schmidt vector $\lambda(\psi) = (\lambda_1, \lambda_2, \lambda_2, \lambda_4)$ for which $\lambda_2 = \lambda_3$. This vertex
fulfills the four inequalities
\begin{align*}
 &E_k^\mathbb{1}(\lambda') \leq E_k(\lambda(\psi)), \quad k \in \{1,2,3\}\\
 &E_2^{\tau_{2,3}}(\lambda') \leq E_2(\lambda(\psi)),
\end{align*}
out of the set $\varOmega$, which defines $M_s^{LU}(\psi)$ (see Eq. (\ref{Ineq_Majunsort})), with equality. Hence, $\lambda(\psi)$ is an element of more than $d-1 = 3$, more precisely of four,
facets which
in turn implies that the three-dimensional polytope is not simple. It is also straightforward to show that $\lambda(\psi)$ has more than $d-1 = 3$ neighbors. Indeed,
each of the following vertices
\begin{align*}
&(\lambda_2,\lambda_2,\lambda_1,\lambda_4),
 (\lambda_1,\lambda_4,\lambda_2,\lambda_2),\\
&(\lambda_2,\lambda_1,\lambda_2,\lambda_4),
 (\lambda_1,\lambda_2,\lambda_4,\lambda_2),
\end{align*}
is an element of $d-2 = 2$ facets that also contain $\lambda(\psi)$ and is thus a neighbor of $\lambda(\psi)$. We therefore conclude that Eq.
(\ref{Eq_Volume}) is not
directly applicable to non-generic states.\\

However, in deriving Eq. (\ref{eq:SourceVolume}) one obtains that
\begin{align}
&V_s(\psi) = \nonumber\\
&\frac{1}{d!} \frac{\sqrt{d}}{(d-1)!} \prod_{k=1}^{d-1} \frac{\abs{\lambda_k-\lambda_{k+1}}}{\lambda_k - \lambda_{k+1}} \sum\limits_{\sigma \in \Sigma_d} \frac{\left( \sum\limits_{k=1}^d \sigma(k) \lambda_k - \frac{d+1}{2}\right)^{d-1}}{\prod\limits_{k=1}^{d-1} \sigma(k) - \sigma(k+1)}.
\label{eq:VolumeFormulaGeneral}
\end{align}
Due to the denominator in the first product of Eq. (\ref{eq:VolumeFormulaGeneral}), degenerate states must be treated with care in deriving a simple
expression for the source volume. 
We show that, for any direction (along non-degenerate states) from which we approach $\lambda(\psi)$ within the subset of sorted Schmidt
vectors, we obtain in the limit $V_s(\psi)$ (given by Eq. (\ref{eq:SourceVolume})) as the volume of the source set. 
Stated differently, for any non-degenerate, sorted Schmidt vector
$\tilde{\lambda}$ and any $\epsilon > 0$ we define the non-degenerate, sorted Schmidt vector
$\lambda(\tilde{\lambda},\epsilon) = (1-\epsilon)\lambda(\psi)+\epsilon \tilde{\lambda}$ for which it holds
that
\begin{align}
\lim_{\epsilon \rightarrow 0} V_s(\lambda(\tilde{\lambda},\epsilon)) = V_s(\psi), \label{eq:Limit}
\end{align}
with $V_s(\psi)$ given by Eq. (\ref{eq:SourceVolume}).
Here, we used that the set of ordered Schmidt vectors is convex and that $\lambda(\tilde{\lambda},\epsilon)$ is a non-degenerate Schmidt vector for which
Eq. (\ref{eq:VolumeFormulaGeneral}) is defined. Let us now show that Eq. (\ref{eq:Limit}) is valid.
We denote by $\mathcal{I}$ the set of all indices $i \in \{1,\ldots,d\}$ for which $\lambda_i = \lambda_{i+1}$. Then we obtain the following.
\begin{widetext}
\begin{align*}
 &\lim\limits_{\epsilon\rightarrow 0}V_s(\lambda(\tilde{\lambda},\epsilon)) = \\
 &= \frac{1}{d!}\frac{\sqrt{d}}{(d-1)!}
 \lim\limits_{\epsilon \rightarrow 0} \left \{\underbrace{\prod_{k \notin \mathcal{I}}  \frac{\abs{(1-\epsilon)(\lambda_k-\lambda_{k+1}) +
 \epsilon (\tilde{\lambda}_k-\tilde{\lambda}_{k+1})}}{(1-\epsilon)(\lambda_k-\lambda_{k+1}) + \epsilon (\tilde{\lambda}_k-\tilde{\lambda}_{k+1})}}_{\rightarrow 1}
 \cdot \underbrace{\prod_{k \in \mathcal{I}} \frac{\abs{\epsilon (\tilde{\lambda}_k - \tilde{\lambda}_{k+1})}}{\epsilon (\tilde{\lambda}_k - \tilde{\lambda}_{k+1})}}_{ \rightarrow 1}
 \cdot \sum\limits_{\sigma \in S_d} \frac{\left( \sum\limits_{k=1}^d \sigma(k)
 \lambda(\tilde{\lambda},\epsilon)_k - \frac{d+1}{2}\right)^{d-1}}{\prod\limits_{k=1}^{d-1} \sigma(k) - \sigma(k+1)} \right \}\\
 & = \frac{1}{d!} \frac{\sqrt{d}}{(d-1)!} \sum\limits_{\sigma \in S_d} \frac{\left( \sum\limits_{k=1}^d \sigma(k) \lambda_k - \frac{d+1}{2}\right)^{d-1}}{\prod\limits_{k=1}^{d-1} \sigma(k) - \sigma(k+1)} = V_s(\psi).
\end{align*}
\end{widetext}
As $\epsilon,\tilde{\lambda}_k - \tilde{\lambda}_{k+1} > 0$ the second product gives 1 and the limit can be computed easily. We thus proved that formula Eq. (\ref{eq:SourceVolume}) for the source volume also holds for degenerate states and
is therefore applicable in general.

An example of a degenerate state is the maximally entangled state $\ket{\phi^+}$ with
Schmidt vector $1/\sqrt{d} \cdot \phi^+$. Since the maximally entangled state cannot be obtained from any other state via LOCC its source
volume has to be zero. Indeed, it holds that
\begin{align*}
&\sum\limits_{k=1}^d \sigma(k) (\phi^+)_k - \frac{d+1}{2} = \frac{1}{d} \sum\limits_{k=1}^d \sigma(k) - \frac{d+1}{2} = 0.
\end{align*}
If one inserts this expression into Eq. (\ref{eq:SourceVolume}) it is easy to see that $V_s(\phi^+) = 0$.

\section{Two different approaches to compute the source and accessible volume}
\label{sec:AppendixB}
In the main part we used two different ways to calculate the $(d-1)$-dimensional volume of convex polytopes in
the space $A \subset \R^d$.
In the first one, used in the derivation of the source volume, we directly determined the volume by considering the $d$-dimensional
Schmidt vectors in this $(d-1)$-dimensional subspace of $\R^d$.  However, in order to determine the accessible volume we considered the projection of the
accessible volume onto
the subspace of $\R^{d}$ that is spanned by the first $(d-1)$ standard vectors. That is, we treated the accessible polytope as a subset of
the set $\{(\lambda_1,\ldots,\lambda_{d-1})| \lambda_i \geq 0, \sum_i \lambda_i \leq 1\} \subset \R^{d-1}$. However, in order to use the same measure
for both, the source and the accessible volume, the volume element in the subspace that contains the projection of the latter
has to be multiplied by the Jacobian of the coordinate transformation
that relates the two methods. As we explain in the following, this amounts to multiplying the volume of the projection of the accessible set by a constant
factor of $\sqrt{d}$.

We have to determine how a volume element expressed
in an orthonormal basis (ONB) of $U = \{\lambda \in \R^d \ \mbox{s.t.} \ \sum_i \lambda_i = 0\}$ is related to a volume element in the projection
on $\R^{d-1}$, i.e. the subspace on which we project the accessible set. It is
straightforward to show that the following vectors are an ONB of $\R^{d}$,
\begin{align*}
 &\mu_0 = \phi^+\\
 &\mu_k = \frac{1}{\sqrt{k}\sqrt{k+1}}(1,\ldots,1,-k,0,\ldots,0) \ \text{for $k \in \{1,\ldots,d-1\}$}
\end{align*}
where in $\mu_k$ a total of $k$ entries of 1 are followed by one entry of $-k$ and zeros. The vectors $\{\mu_k\}_{k=1}^{d-1}$ are also an ONB of $U$. 
If a vector $\lambda \in \R^d$ has coordinates $(u_0,u_1,\ldots,u_{d-1})^T$ in this ONB, its coordinates in the standard basis 
$\vec{\lambda} = (\lambda_1,\ldots,\lambda_{d-1},1-\sum_{i=1}^{d-1}\lambda_i)^T$ are given by 
\begin{align}
\vec{\lambda} = \sum\limits_{k=0}^{d-1} u_k \vec{\mu}_k \label{eq:CoordTrafo},
\end{align}
where $u_0 = 1/\sqrt{d}$ for a normalized Schmidt vector. It is therefore easy to find the $d-1$ Schmidt coefficients $\{\lambda_i\}$ as a function of the
coefficients $\{u_i\}$.
A volume element in $U$ is given by $\prod_{i=1}^{d-1} du_i$, while a volume element in the projected space is given by $\prod_{i=1}^{d-1}d\lambda_{i}$. It
is well-known that these volumes are related by the Jacobian, $\frac{\partial (\lambda_1,\ldots,\lambda_{d-1})}{\partial(u_1,\ldots,u_{d-1})}$, according to
\begin{align}
\prod_{i=1}^{d-1}d\lambda_{i} = \abs{\text{det}\left(\frac{\partial (\lambda_1,\ldots,\lambda_{d-1})}{\partial(u_1,\ldots,u_{d-1})}\right)} \prod_{i=1}^{d-1} du_i.
\end{align}
Using Eq. (\ref{eq:CoordTrafo}) it is easy to show that
\begin{align*}
&\abs{\text{det}\left(\frac{\partial (\lambda_1,\ldots,\lambda_{d-1})}{\partial(u_1,\ldots,u_{d-1})}\right)} = \\
&=\prod\limits_{k=1}^{d-1}\frac{1}{\sqrt{k}\sqrt{k+1}}
\abs{\text{det}\begin{pmatrix} 1& 1& 1& \ldots& 1\\ -1& 1& 1& \ldots&1\\0 &-2 & 1&\ldots&1\\0 &0 & -3&\ldots&1\\ \vdots & & & & \vdots\\ 0 &0 &0 &\ldots &1\end{pmatrix}}\\
&=\frac{1}{\sqrt{d}},
\end{align*}
Hence, we have that $\prod_{i=1}^{d-1}d\lambda_{i} = \frac{1}{\sqrt{d}} \prod_{i=1}^{d-1} du_i$ and the volume of the projection
is $\sqrt{d}$-times smaller than the original volume. However, since this is a constant factor and since we rescale the volumes in order to obtain
the entanglement measures, the two different ways to calculate the volumes of the convex polytopes lead to the same result for the entanglement measures.

\section{Proof of LOCC convertibility conditions}

In \citep{mes} many necessary and sufficient conditions for LOCC transformations of generic four-qubit states were obtained (in particular case (i) and case (iii) have been considered there).  Whereas in \cite{mes} only the conditions for a state to be reachable (or convertible) have been derived, Lemma \ref{Lemma2} gives the necessary and sufficient conditions for the existence of a LOCC-protocol. In order to prove Lemma \ref{Lemma2} we use that LOCC is strictly contained in the set of separable (SEP) operations, for which the criterion for their existence for state transformations has been derived in \cite{gour}. Here, we first review these results and then use them to prove Lemma \ref{Lemma2}.

 Let us denote by $S(\Psi_{seed}) = \lbrace S \in \mathcal{G}: S \ket{\Psi}_{seed} = \ket{\Psi}_{seed} \rbrace$ the set of symmetries of the seed state, 
 where $\mathcal{G}$ denotes the set of local invertible operators. Then for $g, h \in \mathcal{G}$ the state $\ket{\Psi} \propto g \ket{\Psi}_{seed}$ can be transformed into the state $ \ket{\Phi} \propto h\ket{\Psi}_{seed}$ via SEP iff there exists a $m \in \N$, a set of probabilities $\lbrace p_k\rbrace_{k=1}^m$ and $S_k \in S(\Psi_{seed})$ such that \cite{gour}
\begin{align}
\sum_k p_k S_k^{\dagger} H S_k = r G.
\label{sep}
\end{align}
Here, $H = h^{\dagger} h$, $G = g^{\dagger} g$ are local operators and $r = \Vert \ket{\Phi} \Vert^2 / \Vert \ket{\Psi} \Vert^2$.
This criterion is used in \cite{mes} to derive the following necessary conditions for LOCC convertibility of generic four-qubit states.
For these states given in Eq.\ \eqref{standardForm} there are finitely many symmetries of the seed states given by $\lbrace \sigma_i \rbrace_{i=0}^3$. Hence, Eq.\ \eqref{sep} implies \cite{mes}
\begin{equation}
\mathcal{E}_4(H) =\mathcal{E}_1(H^1) \otimes \mathcal{E}_1(H^2) \otimes \mathcal{E}_1(H^3) \otimes \mathcal{E}_1(H^4),
\label{necCond}
\end{equation}
where we used the same notation as in \cite{mes}, i.e. $H = \bigotimes H^i$. Here, $\mathcal{E}_4$ is the completely positive map given by Eq.\ \eqref{sep}, i.e. $\mathcal{E}_4 (H) = \sum_k p_k \sigma_k^{\otimes 4} H \sigma_k^{\otimes 4}$ and $\mathcal{E}_l (\rho) = \sum_k p_k \sigma_k^{\otimes l} \rho \sigma_k^{\otimes l}$. Note that by taking the trace of Eq.\ \eqref{sep} one obtains $ r = 1$.  Furthermore, by tracing over the last two subsystems in Eq.\ \eqref{necCond}, one obtains $\mathcal{E}_2(H^1 \otimes H^2) = \mathcal{E}_1(H^1) \otimes \mathcal{E}_1(H^2)$ (and similarly for other subsystems). Whereas this equation has been used in \cite{mes} to identify the reachable states, here, we mainly use the necessary condition obtained by tracing over all but one system, i.e. 
\begin{equation}
\mathcal{E}_1 (H^1) = G^1,
\label{necCond}
\end{equation}
which is equal to $\vec{\eta} \bigodot \boldsymbol{\zeta}^1 = \boldsymbol{\gamma}^1$.
We use again the same notation as in \cite{mes} with $\bigodot$ denoting the Hadamard product (i.e. entry-wise multiplication), $\vec{\eta} = (\eta_1, \eta_2, \eta_3)^T$, $\eta_0 = \sum_{k=0}^3 p_k = 1$ and $\eta_i = p_0 + p_i - (p_j + p_k)$, with $\lbrace i, j, k \rbrace = \lbrace  1,2,3 \rbrace$.  Hence, we use the necessary and sufficient condition for SEP transformations in Eq.\ \eqref{sep} to obtain necessary conditions for the LOCC convertibility, Eq.\ \eqref{necCond}, and show that they are also sufficient by constructing a corresponding LOCC protocol.

\subsection{Proof of Observation \ref{Observation}}

Due to the fact that the non--isolated generic four--qubit states are of the form $g \ket{\Psi}_{seed}$ \cite{mes}, with $g\in \bigcup_{w\in \{x,y,z\}} {\cal G}_w$, where ${\cal G}_w \equiv \{g^1 \otimes g_w^2 \otimes g_w^3 \otimes g_w^4,g_w^1 \otimes g^2 \otimes g_w^3 \otimes g_w^4,g_w^1 \otimes g_w^2 \otimes g^3 \otimes g_w^4,g_w^1 \otimes g_w^2 \otimes g_w^3 \otimes g^4\}$, we only need to consider transformations among these states.
First we show that $g \ket{\Psi}_{seed}$ with $g \in {\cal G}_w$ can only be transformed to a state $h \in {\cal G}_w$ for the same value of $w$.
This can be easily seen by contradiction. Suppose that $h \in {\cal G}_v$ with $v\neq w$. There always exists an $i \in \{1,2,3,4\}$ for which Eq.\ \eqref{necCond} implies that $\mathcal{E}_1(H_v^i) = \sum_k p_k \sigma_k H_v^i \sigma_k = G_w^i$. However this cannot be fulfilled for $v\neq w$ as $\sigma_k H_v^i \sigma_k \in \text{span}\{\one,\sigma_v\}$ for any $k$, whereas $G_w^i\in \text{span}\{\one,\sigma_w\}$. Hence, the transformation above is only possible if $v = w$. For the remainder of the proofs we will therefore set $w=v=x$ in order to simplify notations. The other cases, i.e. $v=y,z$ can be straightforwardly obtained from that. A similar argument as above shows that the position of the operators which do not only have a component in $x$--direction must coincide. That is, e.g. a transformation of the form
\begin{equation}
g_x^1 \otimes g^2 \otimes g_x^3 \otimes g_x^4 \ket{\Psi}_{seed} \xrightarrow{LOCC} h^1 \otimes h_x^2 \otimes h_x^3 \otimes h_x^4 \ket{\Psi}_{seed}
\end{equation}
is not possible. This completes the proof of Observation \ref{Observation}. \qed \\
Hence, due to Observation \ref{Observation}, the only possible transformations that have to be investigated for LOCC are (up to permutations of the parties)
\begin{equation}
g^1 \otimes g_x^2 \otimes g_x^3 \otimes g_x^4 \ket{\Psi}_{seed} \xrightarrow{LOCC} h^1 \otimes h_x^2 \otimes h_x^3 \otimes h_x^4 \ket{\Psi}_{seed}.
\end{equation}
\subsection{Proof of Lemma \ref{Lemma2}}
Note again that Lemma \ref{Lemma2} [case (ia)] has been proven already in \cite{mes}. Out of convenience we proof Lemma \ref{Lemma2} [case (ib)] at the end of this appendix, as we use some of the results obtained in the proofs of the other cases there.  Hence, we go directly to Lemma \ref{Lemma2} [case (ii)], where transformations of the form (up to permutations)
\begin{align}
g_y^1 \otimes g_x^2 \otimes \mathbb{1} \otimes \mathbb{1} \ket{\Psi}_{seed} \xrightarrow{LOCC} h_y^1 \otimes h_x^2 \otimes \mathbb{1} \otimes \mathbb{1} \ket{\Psi}_{seed}
\end{align}
are considered. Obviously, the proof works analogously if we choose $g_z^1$ instead of $g_y^1$. It follows from Eq.\ \eqref{necCond}, i.e. the condition $\mathcal{E}_1(H_y^1) = G_y^1$, that $\eta_2 \zeta_2^1 = \gamma_2^1$ and thus, $\gamma_2^1$ can only be increased, i.e. $\gamma_2^1 \leq \zeta_2^1$. We determine a simple two-outcome POVM that realizes the transformation. The POVM applied by the first party is given by $\lbrace \sqrt{p} h_y^1 (g_y^1)^{-1}, \sqrt{1-p} h_y^1 \sigma_x (g_y^1)^{-1} \rbrace$ with $p \in [0,1]$.  This is indeed a POVM if $(2p - 1) \zeta_2^1 = \gamma_2^1$ and can be implemented by LOCC as for the second outcome all other parties simply apply the LU $\sigma_x$.  After applying this POVM the state $\ket{\Psi(h_y^1,g_x^1, \mathbb{1}, \mathbb{1})}$ is obtained. The second transformation works analogously. Again, the parameter $\gamma_1^2$ can only be increased by LOCC, i.e. $\gamma_1^2 \leq \zeta_1^2$. The second party applies the POVM $\lbrace \sqrt{p} h_x^2 (g_x^2)^{-1}, \sqrt{1-p} h_x^2 \sigma_y (g_x^2)^{-1} \rbrace$, which can also be implemented by LOCC and thus, Lemma \ref{Lemma2} [case (ii)] is proven. \\
For transformations of the form (Lemma \ref{Lemma2} [case (iiia)])
\begin{equation}
g^1 \otimes \mathbb{1}  \otimes \mathbb{1} \otimes \mathbb{1} \ket{\Psi}_{seed} \rightarrow h^1 \otimes  \mathbb{1} \otimes \mathbb{1} \otimes \mathbb{1} \ket{\Psi}_{seed},
\label{trafo1g}
\end{equation}
 the necessary and sufficient condition for the existence of such a separable transformations can be easily seen to be given by (see Eq. \eqref{necCond})
\begin{equation}
G^1 = \mathcal{E}_1(H^1) = 1/2 \mathbb{1} + \sum_i \zeta_i^1 \eta_i \sigma_i,
\label{nec1g}
\end{equation}
where as defined before $\eta_0 = \sum_{k=0}^3 p_k = 1$ and $\eta_i = p_0 + p_i - (p_j + p_k)$, with $\lbrace i, j, k \rbrace = \lbrace  1,2,3 \rbrace$ and $p_i\geq 0$ for any $i$. Due to the condition above we have for $\zeta^1_i\neq 0$ for all $i$, $p_i = 1/4(1 +r_i - r_j - r_k)$ for $r_i = \gamma_i^1 / \zeta_i^1$, where $\lbrace i, j, k \rbrace = \lbrace  1,2,3 \rbrace$ and $p_0 = 1/4(1 + r_1 + r_2 + r_3)$. Hence the necessary and sufficient conditions for the existence of a separable transformation are that all these probabilities are non--negative, as stated in Lemma \ref{Lemma2} [case (iiia)]. Note that any of these separable transformations can be realized via LOCC as the POVM $\lbrace M_i\rbrace_{i=1}^4 $ \cite{mes} with $M_i = \sqrt{p_i} h^1 \sigma_i (g^1)^{-1} \otimes (\sigma_i)^{\otimes 3}$ can be implemented locally. Hence, if $\zeta^1_i\neq 0$ for all $i$ the necessary and sufficient conditions for the existence of the LOCC transformation are  $1+ r_i - r_j - r_k \geq 0$ for any choice of $i,j,k$ such that $\{i,j,k\}=\{1,2,3\}$.

Note that we have different necessary and sufficient conditions for transformations as in Eq.\ \eqref{trafo1g} if the state $\ket{\Phi(h^1,\mathbb{1}, \mathbb{1}, \mathbb{1})}$ has vanishing $\zeta_i^1$ parameters as then $r_i$ above is not defined. For states with one vanishing parameter (Lemma\ \ref{Lemma2} [case (iiib)], i.e. without loss of generality $\zeta_1^1=0$, Eq.\ \eqref{nec1g} implies that also the corresponding $\gamma_i^1$ parameter has to be equal to zero, i.e. $\gamma_1^1 =0$. The other two parameters have to fulfill $\eta_{2(3)} \zeta_{2(3)}^{1} = \gamma_{2(3)}^1$ with $\eta_i = p_0 + p_i - (p_j + p_k)$, $\lbrace i,j,k \rbrace = \lbrace 1,2,3 \rbrace$. As $\eta_2$ and $\eta_3$ are linearly independent and the two parameters $\gamma_2^1$, $\gamma_3^1$ can be chosen nonnegative in the standard form (Eq.\ \eqref{standardForm}) the state $\ket{\Phi(h^1,\mathbb{1}, \mathbb{1}, \mathbb{1})}$ can be reached from all states $\ket{\Psi(g^1,\mathbb{1}, \mathbb{1}, \mathbb{1})}$ with $\gamma_2^1 \leq \zeta_2^1$ and $\gamma_3^1 \leq \zeta_3^1$.
A similar condition holds for states $\ket{\Phi(h^1,\mathbb{1}, \mathbb{1}, \mathbb{1})}$  with two vanishing parameters, i.e. without loss of generality $\zeta_1^1= \zeta_2^1 = 0$. Then, from Eq.\ \eqref{nec1g} it follows again that the corresponding $\gamma_i^1$ parameters are equal to zero, i.e. $\gamma_1^1 = \gamma_2^1 = 0$. Furthermore, the condition $\eta_3 \zeta_3^1 = \gamma_3^1$ has to be fulfilled, thus, the only necessary and sufficient conditions for transforming $\ket{\Psi(g_z^1,\mathbb{1}, \mathbb{1}, \mathbb{1})}$ into $\ket{\Phi(h_z^1,\mathbb{1}, \mathbb{1}, \mathbb{1})}$ is $\gamma_3^1 \leq \zeta_3^1$. \\
Note that the necessary conditions for reaching the state $\ket{\Phi(h_x^1, h_{y,z}^2,\mathbb{1},\mathbb{1})}$ via LOCC (Lemma \ref{Lemma2} [case (ib)]) can be obtained by considering the equation $\mathcal{E}_2 (H_x^1 \otimes H_{y,z}^2) = \mathcal{E}_1(H_x^1) \otimes \mathcal{E}_1(H_{y,z}^2)$ (see \cite{mes}). From there it follows that $\eta_1 \eta_2 - \eta_3 = 0$ and $ \eta_1 \eta_3 - \eta_2 = 0$. Hence, these conditions are on the one hand fulfilled if $(\eta_1)^2 = 1$, which leads to the same necessary and sufficient conditions as in Lemma \ref{Lemma2} [case (ia)]. The states fulfilling these conditions are of the form $\ket{\Psi(h_x^1, g_{y,z}^1, \mathbb{1},\mathbb{1})}$  and a POVM allowing for this transformation via LOCC is given in \cite{mes}. On the other hand the necessary conditions are also fulfilled if $\eta_2 = \eta_3 = 0$. By taking into account Eq.\ \eqref{necCond} this implies that $\eta_1 \zeta_1^1 = \gamma_1^1$ and $\gamma_2^2 = \gamma_3^2 = 0$. Thus, $\ket{\Phi(h_x^1, h_{y,z}^2,\mathbb{1},\mathbb{1})}$ can also be reached by states of the form $\ket{\Psi(g_x^1, \mathbb{1},\mathbb{1},\mathbb{1})}$ with $\gamma_1^1 \leq \zeta_1^1$. The POVM realizing this transformation via LOCC are already given in \cite{mes} and in the proof of Lemma \ref{Lemma2} [case (iii)] above, as we simply first transform $\ket{\Psi(g_x^1, \mathbb{1},\mathbb{1},\mathbb{1})}$ into $\ket{\Phi(h_x^1, \mathbb{1},\mathbb{1},\mathbb{1})}$ and then the latter state is converted into the final state $\ket{\Phi(h_x^1, h_{y,z}^2,\mathbb{1},\mathbb{1})}$.
\qed \\

\end{appendix}

\end{document}